\documentclass[journal]{IEEEtran}
\ifCLASSINFOpdf
\else
\fi
\usepackage{cite}
\usepackage{amsmath}
\usepackage[pdftex]{graphicx} 
\usepackage{subfig}
\usepackage{array} 

\usepackage{booktabs} 
\usepackage{amssymb}
\usepackage{mathtools}
\usepackage{cancel} 
\usepackage{bm} 
\usepackage{esint} 
\usepackage{amsthm}
\usepackage{tikz}
\usepackage{tikz-3dplot}
\usetikzlibrary{backgrounds,3d,shadings,shapes.misc,decorations.pathmorphing,shapes,calc,fadings,shadows.blur}

\usepackage{refcount}
\usepackage{xfrac}

\usepackage{amssymb,amsmath}

\usepackage{bm}
\newcommand{\bs}[1]{\bm{\mathrm{#1}}}

\newcommand{\vect}[1]{\vec{#1}}
\newcommand{\nhat}{\ensuremath{\hat{n}}}

\newcommand{\vecprime}[1]{\vect{#1}^{\,\prime}}

\renewcommand{\r}{\left(\vect{r}\right)}
\newcommand{\rp}{\left(\vecprime{r}\right)}

\newcommand{\rrpk}[1][]{\left(k_{#1}, \vect{r}, \vecprime{r}\right)}
\newcommand{\rrpki}[1][]{\left(k_i, \vect{r}, \vecprime{r}\right)}

\newcommand{\rin}[1]{\left(\vect{r}\right)\big\rvert_{#1}}
\newcommand{\matr}[1]{\bs{#1}}

\newcommand{\abs}[1]{\left \lvert #1 \right\rvert }

\newcommand{\pvint}{\dashint}

\newcommand{\junk}[1] {}

\def\Xint#1{\mathchoice
{\XXint\displaystyle\textstyle{#1}}%
{\XXint\textstyle\scriptstyle{#1}}%
{\XXint\scriptstyle\scriptscriptstyle{#1}}%
{\XXint\scriptscriptstyle\scriptscriptstyle{#1}}%
\!\int}
\def\XXint#1#2#3{{\setbox0=\hbox{$#1{#2#3}{\int}$}
\vcenter{\hbox{$#2#3$}}\kern-.5\wd0}}

\def\dashint{\Xint-}

\newcommand*\widebar[1]{%
  \hbox{%
    \vbox{%
      \hrule height 0.5pt 
      \kern0.3ex
      \hbox{%
        \kern-0.05em
        \ensuremath{#1}%
        \kern-0.05em
      }%
    }%
  }%
} 

\newcommand{\mathsout}[1]
{\bgroup\mathchoice
  {\sbox0{$\displaystyle{#1}$}%
    \usebox0\hspace{-\wd0}%
    \rule[0.5\ht0-0.2\dp0-.5pt]{.7\wd0}{0.8pt}\hspace{.3\wd0}}%
  {\sbox0{$\textstyle{#1}$}%
    \usebox0\hspace{-\wd0}%
    \rule[0.5\ht0-0.2\dp0-.5pt]{.7\wd0}{0.8pt}\hspace{.3\wd0}}%
  {\sbox0{$\scriptstyle{#1}$}%
    \usebox0\hspace{-\wd0}%
    \rule[0.5\ht0-0.2\dp0-.5pt]{.7\wd0}{0.8pt}\hspace{.3\wd0}}%
  {\sbox0{$\scriptscriptstyle{#1}$}%
    \usebox0\hspace{-\wd0}%
    \rule[0.5\ht0-0.2\dp0-.5pt]{.7\wd0}{0.8pt}\hspace{.3\wd0}}%
\egroup}

\renewcommand{\epsilon}{\varepsilon}

\newcommand{\opL}[1][]{\ensuremath{\mathcal{L}_{#1}}} 
\newcommand{\opK}[1][]{\ensuremath{\mathcal{K}_{#1}}} 
\newcommand{\opKpv}[1][]{\mathsout{\opK[#1]}} 
\newcommand{\opM}[1][]{\ensuremath{\mathcal{M}_{#1}}} 
\newcommand{\opMd}[1][]{\ensuremath{\mathcal{M}^{\dagger}_{#1}}}
\newcommand{\opMpv}[1][]{\mathsout{\opM[#1]}} 
\newcommand{\opMdpv}[1][]{\mathsout{\opMd[#1]}}

\newcommand{\Lmat}[1][]{{\matr{L}_{#1}}}

\newcommand{\Kpvmat}[1][]{{\mathsout{\matr{K}}_{#1}}}

\newcommand{\Mpvmat}[1][]{{\mathsout{\matr{M}}_{#1}}}

\renewcommand{\tt}{{(\mathrm{tt})}}

\newcommand{\Lttmat}[1][]{{\matr{L}_{#1}^{\mathrm{(ff)}}}}
\newcommand{\Lssmat}[1][]{{\matr{L}_{#1}^{\mathrm{(hh)}}}}

\newcommand{\Lnnmat}[1][]{{\matr{L}_{#1}^{\mathrm{(hh)}}}}
\newcommand{\Lntmat}[1][]{{\matr{L}_{#1}^{\mathrm{(hf)}}}}
\newcommand{\Ltnmat}[1][]{{\matr{L}_{#1}^{\mathrm{(fh)}}}}

\newcommand{\Msnmat}[1][]{{\matr{M}_{#1}^{\mathrm{(hh)}}}}
\newcommand{\Kttmat}[1][]{{\matr{K}_{#1}^{\mathrm{(fg)}}}}

\newcommand{\Mnspvmat}[1][]{{\mathsout{\matr{M}}_{#1}^{\mathrm{(hh)}}}}
\newcommand{\Msnpvmat}[1][]{{\mathsout{\matr{M}}_{#1}^{\mathrm{(hh)}}}}
\newcommand{\Kttpvmat}[1][]{{\mathsout{\matr{K}}_{#1}^{\mathrm{(fg)}}}}

\newcommand{\Kntpvmat}[1][]{{\mathsout{\matr{K}}_{#1}^{\mathrm{(hg)}}}}

\newcommand{\Irtmat}{{\matr{G}^{\mathrm{(fg)}}}}
\newcommand{\Issmat}{{\matr{G}^{\mathrm{(hh)}}}}

\newcommand{\figref}[1]{Fig.~\ref{#1}}
\newcommand{\secref}[1]{Section~\ref{#1}}

\usepackage{tcolorbox}

\usepackage{textpos}

\newcommand{\mySubtitle}[1]%
{%
	\begin{textblock}{14.0}(0.7, 2.9)
		\textbf{#1}%
	\end{textblock}%
}%

\newcommand{\redcol}{black}

\newcommand{\red}[1]{\textcolor{\redcol}{#1}}
\newcommand{\green}[1]{\textcolor{green!0!black}{#1}}

\newcommand{\brown}[1]{\textcolor{black}{#1}}
\newcommand{\magenta}[1]{\textcolor{black}{#1}}

\newcommand{\Ecolor}[1]{{#1}}
\newcommand{\Hcolor}[1]{{#1}}
\newcommand{\Jcolor}[1]{{#1}}

\newcommand{\Rhocolor}[1]{\brown{#1}}

\newcommand{\Er}[1][]{\Ecolor{\vect{E}_{#1}\r}}

\newcommand{\Hr}[1][]{\Hcolor{\vect{H}_{#1}\r}}

\newcommand{\Htr}[1][]{\Hcolor{\nhat_{#1} \times \vect{H}_{#1}\r}}

\newcommand{\Jr}[1][]{\ensuremath{\Jcolor{\vect{J}_{#1}\r}}}

\newcommand{\Jmat}[1][]{\ensuremath{\Jcolor{\matr{J}_{#1}}}}

\newcommand{\Grrpk}[1][]{\ensuremath{\green{G_{#1}\rrpk}}}

\newcommand{\Grrpki}[1][]{\ensuremath{\green{G\rrpki}}}

\newcommand{\gradrGrrpk}[1][]{\ensuremath{\green{\nabla G_{#1}\rrpk}}}

\newcommand{\Ar}[1][]{\magenta{\vect{A}_{#1}\r}}

\newcommand{\Atr}[1][]{\magenta{\nhat \times \vect{A}_{#1}\r}}
\newcommand{\Atrp}[1][]{\magenta{\nhat' \times \vect{A}_{#1}\rp}}

\newcommand{\Anr}[1][]{\magenta{\nhat \cdot \vect{A}_{#1}\r}}
\newcommand{\Anrp}[1][]{\magenta{\nhat' \cdot \vect{A}_{#1}\rp}}

\newcommand{\divrAr}[1][]{\magenta{\nabla \cdot \vect{A}_{#1}\r}}
\newcommand{\divrpArp}[1][]{\magenta{\nabla' \cdot \vect{A}_{#1}\rp}}

\newcommand{\curlrAr}[1][]{\magenta{\nabla \times \vect{A}_{#1}\r}}
\newcommand{\curlrpArp}[1][]{\magenta{\nabla' \times \vect{A}_{#1}\rp}}

\newcommand{\rhor}[1][]{\red{\rho_{#1}\r}}

\newcommand{\Phimat}[1][]{\magenta{\matr{\Phi}_{#1}}}

\newcommand{\phir}[1][]{\magenta{\phi_{#1}\r}}
\newcommand{\phirp}[1][]{\magenta{\phi_{#1}\rp}}

\newcommand{\gradrphir}[1][]{\magenta{\nabla\phi_{#1}\r}}

\newcommand{\Apr}[1][]{\magenta{\vect{A}_{#1}\,'\r}}
\newcommand{\phipr}[1][]{\magenta{\phi_{#1}'\r}}
\newcommand{\gradrphipr}[1][]{\magenta{\nabla\phi_{#1}'\r}}

\newcommand{\chir}[1][]{\magenta{\chi_{#1}\r}}

\newcommand{\phirin}[2][]{\magenta{\phi_{#1}\rin{#2}}}

\newcommand{\gradrphirin}[2][]{\magenta{\nabla\phi_{#1}\rin{#2}}}
\newcommand{\chirin}[2][]{\magenta{\chi_{#1}\rin{#2}}}
\newcommand{\Arin}[2][]{\magenta{\vect{A}_{#1}\rin{#2}}}
\newcommand{\Atrin}[2][]{\magenta{\nhat \times \vect{A}_{#1}\rin{#2}}}

\newcommand{\Anprin}[2][]{\magenta{\nhat \cdot \vect{A}_{#1}\,'\rin{#2}}}
\newcommand{\curlrArin}[2][]{\magenta{\nabla \times \vect{A}_{#1}\rin{#2}}}
\newcommand{\divrArin}[2][]{\magenta{\nabla \cdot \vect{A}_{#1}\rin{#2}}}

\newcommand{\Htrin}[2][]{\Hcolor{\nhat \times \vect{H}_{#1}\rin{#2}}}

\newcommand{\uamat}[1][]{\ensuremath{\Jcolor{\matr{A}_{c,#1}}}}
\newcommand{\ubmat}[1][]{\ensuremath{\Ecolor{\matr{A}_{t,#1}}}}

\newcommand{\udmat}[1][]{\ensuremath{\Rhocolor{\matr{A}_{n,#1}}}}

\makeatletter
\newlength\numerator@height
\newlength\frac@height
\newsavebox\numerator@box
\newsavebox\frac@box

\newcommand\dfracparens[3]{%
	\sbox{\numerator@box}{\ensuremath{#1}}%
	\sbox{\frac@box}{\ensuremath{\dfrac{#1}{#2}}}%
	\settoheight{\frac@height}{\usebox{\frac@box}}%
	\settoheight{\numerator@height}{\usebox{\numerator@box}}%
	\addtolength{\frac@height}{-\numerator@height}%
	\usebox{\frac@box}%
	\raisebox{\frac@height}{%
		\( \left( {#3} \right)
		\)}%
}
\makeatother

\usepackage{dsfont}
\newcommand{\1}[1][]{\mathds{1}_{#1}}

\begin{document}
%
%
\title{Electromagnetic Modeling of Lossy Interconnects From DC to High Frequencies With a Potential-Based~Boundary~Element~Formulation}

%
%
%

\author{Shashwat~Sharma,~\IEEEmembership{Graduate Student Member,~IEEE,}
        and~Piero~Triverio,~\IEEEmembership{Senior Member,~IEEE}
\thanks{S. Sharma is with the Edward S. Rogers Sr. Department of Electrical \& Computer Engineering, University of Toronto, Toronto, ON, M5S 3G4 Canada, e-mail: shash.sharma@mail.utoronto.ca.
P. Triverio is with the Edward S. Rogers Sr. Department of Electrical \& Computer Engineering and with the Institute of Biomedical Engineering, University of Toronto, Toronto, ON, M5S 3G4 Canada, email: piero.triverio@utoronto.ca.}
\thanks{This work was supported by Advanced Micro Devices, by the Natural Sciences and Engineering Research Council of Canada (Collaborative Research and Development Grants program), and by CMC Microsystems.}
\thanks{Manuscript received $\ldots$; revised $\ldots$.}}

%
%

\markboth{IEEE Transactions on Microwave Theory and Techniques}%
{Sharma \MakeLowercase{\textit{et al.}}: Potential-Based EM Modeling of Interconnects}
%



\maketitle

\begin{abstract}
	The accurate electromagnetic modeling of both low- and high-frequency physics is crucial in the signal and power integrity analysis of electrical interconnects.
	The boundary element method (BEM) is appealing for lossy conductor modeling because it can capture the frequency-dependent variation of skin depth with only a surface-based discretization of the structure.
	Conventional BEM formulations rely on the mutual coupling of electric and magnetic fields, and can become inaccurate or unstable at low frequencies.
	We develop a new full-wave BEM formulation based on potentials which can accurately model lossy conductors from exactly DC to very high frequencies.
	A new set of simple boundary conditions is proposed along with a modified Lorenz gauge to ensure that the proposed formulation has a stable condition number down to DC.
	Moreover, coupling the potential-based integral equations to a circuit model allows the straightforward extraction of network parameters.
	Realistic numerical examples at both the chip and package level demonstrate the accuracy and stability of the proposed method from DC to high frequencies, beyond the capabilities of state-of-the-art BEM formulations based on fields.
\end{abstract}


%
\IEEEpeerreviewmaketitle


\section{Introduction}

\IEEEPARstart{A}{dvances} in the design of integrated circuits at both the chip and the package level have made full-wave electromagnetic simulation tools indispensable.
Signal and power integrity analysis requires predicting port parameters such as reflection, transmission, and crosstalk over a wide range of frequency, including both the static and high-frequency limits.
These parameters are heavily influenced by the skin depth, which undergoes large variations over such a wide range of frequency.
Therefore, the skin effect must be modeled accurately from DC to tens or hundreds of gigahertz in typical chip- and package-level applications.

The finite element method~\cite{FEMJin,femDC01DJ,femDC02DJ,femDC03DJ} and volume integral equations~\cite{jandh_peec_2007,PEEC02,PEEC01,VolIE01,fastmaxwell} tend to be robust.
However, they become computationally expensive at high frequencies, where an increasingly fine 3D mesh is needed near the surface of conductors to resolve the shrinking skin depth.
The boundary element method~(BEM)~\cite{book:colton,ChewWAF}, in contrast, utilizes a surface integral representation of Maxwell's equations~\cite{ChewIEM} and requires only a 2D mesh on the surface of conductors, while still capturing the variations in skin depth and the coupling between objects.

A multitude of BEM formulations for full-wave skin effect modeling have been proposed in the literature~\cite{Song2003,gibc,gibcHmatDanJiao,DSA01,DSA07,DSA08,UTK_MTT,EPEPS2017}, all of which compute the tangential electric and/or magnetic field on the surface of each conductive object.
These formulations rely on the coupling between electric and magnetic fields.
At very low frequencies, the fields begin to decouple, which can cause field-based BEM formulations to become poorly conditioned or inaccurate~\cite{lfbreakdown,lfefieDJ,ChewAEFIEperturbOrig}.
Therefore, different computational methods are often needed for different frequency ranges, which can be inconvenient.
For example, surface impedance boundary conditions~\cite{SIBC} are useful for capturing the extremely small skin depth at high frequencies in both boundary element and finite element contexts.
Instead, quasistatic approximations are often used for capacitance~\cite{cap02}, inductance~\cite{fasthenry}, and impedance~\cite{MQSOKH2,MQSOKH,MQSOKH3,UTK_SPI2017} extraction at low frequencies.
Full-wave solvers involving a volumetric mesh for conductors can also be used in the low frequency range~\cite{fastimp,fastmaxwell,PEEC02,PEEC01,VolIE01,FEMJin}, but can struggle to resolve the skin depth at high frequencies.
Combining results from these different tools can lead to discontinuities in the port parameters, which in turn may cause numerical issues and causality violations when computing the time domain response~\cite{TriverioCausality01,TriverioCausality02}.
Moreover, it is not easy to predict the frequency range over which a particular method or formulation is valid, especially for the multiscale structures encountered in chip- and package-level applications: at the same frequency, some geometric features may be operating in the quasistatic regime while others may already be experiencing wave effects.

Several BEM techniques have been proposed to enable accurate electromagnetic modeling at extremely low frequencies.
Loop-star and loop-tree basis functions have been studied extensively~\cite{loopstar_vecchi_1999,loopstar_chew_2000,loopstar_jflee_2003,loopstar_eibert_2004,loopstar_andriulli_2012}, but require searching the mesh for global loops, which can be expensive for complex geometries.
The Calder\'on multiplicative preconditioner was shown to be very effective for perfect conductors~\cite{calderonEFIE,RFCMP02}, conductors with a simple impedance boundary condition~\cite{calderonHPIBC}, and dielectrics~\cite{calderonPMCHWT,calderonSSdiel,DSA_Calderon_HDC}.
Projectors based on a Helmholtz decomposition have also been developed for the case of perfect conductors~\cite{RFCMP03}.
However, the effectiveness of these techniques in the context of chip- and package-level structures is not clear; in some cases, the Calder\'on-preconditioned system of equations seems to become inaccurate for lossy objects~\cite{PIE09}.
Wideband field-based BEM formulations specifically designed for lossy conductor modeling have also been proposed~\cite{agibc,eaefie02,AWPLSLIM,JMMCTSLDM}, but they eventually become inaccurate in the static limit due to the decoupling of the fields.
Indeed, it will be shown in \secref{sec:results} that state-of-the-art BEM formulations based on fields~\cite{agibc,eaefie02} yield inaccurate network parameters below the~${\sim}100\,$kHz--$1\,$MHz range for chip-level structures.

The need for broadband electromagnetic solvers recently sparked an interest in BEM formulations based on electromagnetic potentials, rather than fields~\cite{PIE04,PIE01,PIE03}.
Potential-based integral equations (PIEs) are appealing because they do not exhibit the low-frequency breakdown behavior of field-based methods~\cite{ChewIEM,PIE04,PIE01}.
However, much of the recent literature on PIE methods focuses on perfect electric conductors~\cite{PIE04,PIE03,PIE08,PIE_TD_PEC}, making these formulations unsuitable for chip- and package-level simulations.
Modeling lossy conductors with a full-wave PIE formulation is significantly more challenging than the perfect conductor case because an additional set of unknown quantities and integral operators is required~\cite{PIE01,PIE02}.
Boundary conditions on the potentials must be imposed carefully to ensure that unique solutions are found.
The lossy case was briefly considered in~\cite{PIE09}, but without a mathematical description and only in the context of scattering.

Magneto-quasistatic PIE methods were extensively studied for eddy current modeling in lossy conductors~\cite{eddy01,eddy02,eddy03,eddy04,eddy05}, but are not applicable at high frequencies.
A PIE approach was proposed for dielectrics~\cite{PIE02}, but was studied only in a theoretical sense and may not be suitable for lossy conductors; it requires adding together integral equations associated with adjacent materials, with can be inaccurate for conductors embedded in a dielectric.
A full-wave PIE formulation for lossy conductors was recently proposed~\cite{AWPLVPIE} for electromagnetic scattering analysis.
Though the method in~\cite{AWPLVPIE} is accurate over extremely wide ranges of frequency and conductivity, it requires solving for a relatively large number of unknowns compared to field-based formulations, and does not address port parameter extraction in coupled electromagnetic-circuit problems.

In this article, we propose a new full-wave PIE formulation for electromagnetic problems involving lossy conductors.
The proposed method can be used to extract port parameters for coupled electromagnetic-circuit systems, and is accurate from exactly DC to tens or hundreds of gigahertz for typical interconnect structures at the chip and package levels.
To the best of our knowledge, this DC-to-high-frequency modeling capability has not been achieved by existing BEM formulations for lossy conductors.
We leverage the gauge invariance of potentials~\cite{jacksonEM} to devise new boundary conditions for the vector potential which are simpler than those used in existing PIE methods~\cite{PIE01,PIE02,AWPLVPIE}.
We also discuss a modified form of the Lorenz gauge which allows modeling the magnetic vector potential accurately inside each object, regardless of the choice of reference for the electric scalar potential.
We describe how the PIEs associated with the regions external and internal to each object can be directly coupled to a circuit with the help of the continuity equation and Kirchoff's voltage law (KVL), allowing a straightforward extraction of port parameters.
With realistic numerical examples, we demonstrate that our method can solve coupled electromagnetic-circuit systems at both the chip and package levels from DC to high frequencies.

The goal of this work is to describe the various mathematical considerations involved in developing the proposed formulation, and to study its accuracy for representative examples.
Practical considerations such as incorporating acceleration algorithms and modeling layered substrates are not considered here.
The proposed method is described in \secref{sec:method}, starting with a discussion of the proposed boundary conditions (\secref{sec:bcs}) and the modified Lorenz gauge (\secref{sec:gauge}), followed by a derivation of the pertinent PIEs (\secref{sec:vpie}) and the proposed discretization scheme (\secref{sec:discr}).
Finally, several numerical examples are presented in \secref{sec:results}, followed by concluding remarks in \secref{sec:concl}.

\section{Proposed Formulation}\label{sec:method}

\begin{figure}[t]
	\centering
	\includegraphics[width=0.5\linewidth]{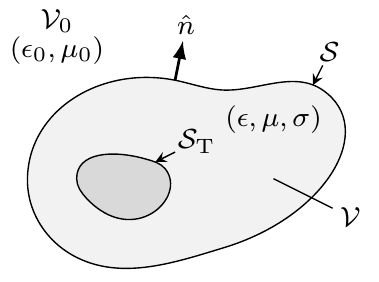}
	\caption{Geometry of objects considered in this work.}\label{fig:geom}
\end{figure}

Consider an object occupying volume~$\mathcal{V}$ with permittivity~$\epsilon$, permeability~$\mu$, and conductivity~${\sigma > 0}$, as shown in \figref{fig:geom}.
Region~$\mathcal{V}$ is bounded by the surface~$\mathcal{S}$ with outward unit normal vector~$\nhat$.
Surface~$\mathcal{S}$ is composed of two parts: $\mathcal{S}_{\mathrm{T}}$ is an electrically small portion of~$\mathcal{S}$ which facilitates the connection to a terminal of an external circuit, while the remaining portion $\mathcal{S}_{\mathrm{U}}$ is not connected.
The object may be attached to multiple terminals, in which case $\mathcal{S}_{\mathrm{T}} = \bigcup_i^{N_\mathrm{T}}\mathcal{S}_{\mathrm{T}i}$, where $\mathcal{S}_{\mathrm{T}i}$ corresponds to the~$i$th terminal surface, and~$N_\mathrm{T}$ is the number of terminals on the object.
The object resides in free space,~$\mathcal{V}_0$, with permittivity~$\epsilon_0$ and permeability~$\mu_0$.
We assume that the attached circuit provides the excitation, and there are no other sources in~$\mathcal{V}$ or~$\mathcal{V}_0$.

For~${\vect{r}\in\mathcal{V}_0}$, the magnetic vector and electric scalar potentials,~$\Apr$ and~$\phipr$, respectively, can be defined via~\cite{jacksonEM}
\begin{align}
	\mu_0\Hr &= \nabla\times\Apr, \label{eq:Adef}\\
	\Er &= -j\omega\Apr - \gradrphipr, \label{eq:phidef}
\end{align}
where~$\omega$ is the angular frequency,~$\Hr$ is the magnetic field, and~$\Er$ is the electric field.
Equations~\eqref{eq:Adef} and~\eqref{eq:phidef} imply that a new set of potentials~$\Ar$ and~$\phir$ can be chosen via the gauge transformation
\begin{align}
	\Ar &= \Apr + \nabla\chir,\label{eq:Atransform}\\
	\phir &= \phipr + j\omega\chir,\label{eq:phitransform}
\end{align}
which leaves~$\Er$ and~$\Hr$ unchanged~\cite{jacksonEM}.
In this work, the scalar function~$\chir$ will be chosen strategically to derive surface integral equations in terms of~$\Ar$ and~$\phir$ with simple boundary conditions.
One integral equation will be devised to model the distribution of potentials within the object~$\mathcal{V}$ (internal problem), and another will be used for~$\mathcal{V}_0$ to capture the coupling between different objects (external problem).

\subsection{Boundary Conditions}\label{sec:bcs}

A key consideration for obtaining potential-based integral equations is the choice of boundary conditions on~$\mathcal{S}$.
To ensure that the scalar potential~$\phipr$ is continuous across~$\mathcal{S}$, we take
\begin{align}
	\phirin{\mathcal{S}^-} &= \phirin{\mathcal{S}^+},\label{eq:bcphir}\\
	\chirin{\mathcal{S}^-} &= \chirin{\mathcal{S}^+},\label{eq:bcchir}
\end{align}
where~$\mathcal{S}^-$ and~$\mathcal{S}^+$ denote the inner and outer sides of~$\mathcal{S}$, respectively, and the reference for~$\phir$ is chosen at infinity.
For the vector potential~$\Ar$, we consider the normal and tangential components separately.
\subsubsection{Normal component of~$\Ar$}
We use the degree of freedom provided by~$\chir$ to simplify the surface integral equations to be derived, by picking~$\chir$ such that
\begin{align}
	\Anr = 0,\quad\left(\vect{r}\in\mathcal{S}^+\right),\label{eq:bcAnr}
\end{align}
which, because of~\eqref{eq:Atransform}, requires that
\begin{align}
	\nhat\cdot\nabla\chirin{\mathcal{S}^+} = -\Anprin{\mathcal{S}^+}.\label{eq:bcnchir}
\end{align}
The existence of a~$\chir$ such that~\eqref{eq:bcAnr} and~$\eqref{eq:bcnchir}$ are satisfied is discussed in the appendix.
Equation~\eqref{eq:bcAnr} also ensures that the neutrality condition~\cite{PIE04,PIE01}
\begin{align}
	\int_{\mathcal{S}^+}d\mathcal{S}\,\Anr = 0\label{eq:Anr0neutr}
\end{align}
is automatically satisfied, which relates to the enforcement of charge neutrality on~$\mathcal{S}^+$~\cite{PIE02}.
The boundary condition~\eqref{eq:bcAnr} is a significant point of departure from existing PIE formulations for penetrable objects~\cite{PIE02,AWPLVPIE}.
In~\cite{AWPLVPIE}, the boundary condition for~$\Anr$ was derived from Gauss' Law and the continuity equation, and requires involving~$\nhat\cdot\gradrphir$ as an additional unknown.
In turn, this requires solving the scalar potential integral equation~\cite{PIE01} in both~$\mathcal{V}$ and~$\mathcal{V}_0$~\cite{AWPLVPIE}.
In contrast, using~\eqref{eq:bcAnr} avoids the need to take~$\nhat\cdot\gradrphir$ as unknown and to solve the scalar potential integral equation.
To obtain a physical interpretation of~\eqref{eq:bcAnr}, we consider for a moment the case when~$\mathcal{V}$ is a perfect conductor.
Then, quantities~$\Anr$ and~$\nhat\cdot\gradrphir$ on~$\mathcal{S}^+$ are contributions towards the surface charge density~$\rhor[\mathrm{s}]$ on~$\mathcal{S}^+$~\cite{PIE01},
\begin{align}
	\epsilon_0^{-1}\rhor[\mathrm{s}] = -j\omega\,\Anr - \nhat\cdot\gradrphir,\quad\left(\vect{r}\in\mathcal{S}^+\right).\label{eq:rhorpec}
\end{align}
Using~\eqref{eq:bcAnr} in~\eqref{eq:rhorpec} gives
\begin{align}
	\epsilon_0^{-1}\rhor[\mathrm{s}] = -\nhat\cdot\gradrphir,\quad\left(\vect{r}\in\mathcal{S}^+\right),\label{eq:rhorpecbc}
\end{align}
valid at any frequency.
Equation~\eqref{eq:rhorpecbc} is rather intuitive because it implies that the electric surface charge density on~$\mathcal{S}^+$ for a perfect conductor is related only to the scalar potential, and not to the vector potential, when gauge freedom~\cite{jacksonEM} is exploited via the proposed boundary condition~\eqref{eq:bcAnr}.
In the formulation we propose,~${\nhat\cdot\gradrphir}$ is not taken as an unknown, but if needed it can be computed as a simple post-processing step by solving the scalar potential integral equation~\cite{PIE01,PIE02}, after~$\phir$ has been computed~\cite{PIE07,PIE08,AWPLVPIE}.
Then,~$\rhor[\mathrm{s}]$ can obtained via~\eqref{eq:rhorpecbc} if desired.
In the case of a lossy conductor,~$\rhor[\mathrm{s}]$ would represent an equivalent surface charge density on~$\mathcal{S}^+$~\cite{PIE01}.

\subsubsection{Tangential component of~$\Ar$}
For the tangential fields~$\Atr$ and~$\nhat\times\curlrAr$, we use the conventional boundary conditions~\cite{PIE01,PIE02}
\begin{align}
	\Atrin{\mathcal{S}^-} = \Atrin{\mathcal{S}^+}\label{eq:bcAtr}
\end{align}
and
\begin{align}
	\frac{1}{\mu_0}\,\nhat\times\curlrArin{\mathcal{S}^-} = \frac{1}{\mu}\,\nhat\times\curlrArin{\mathcal{S}^+}.\label{eq:bcAcr}
\end{align}
Equation~\eqref{eq:bcAcr} follows from~\eqref{eq:Adef},~\eqref{eq:Atransform}, and the continuity of~$\Htr$ across~$\mathcal{S}$,
\begin{align}
	\Htrin{\mathcal{S}^-} = \Htrin{\mathcal{S}^+}.\label{eq:bcHtr}
\end{align}

\subsection{Choice of Gauge}\label{sec:gauge}

Maxwell's equations~\cite{EMharrington} along with~\eqref{eq:Adef}--\eqref{eq:phitransform} can be used to derive a partial differential equation for~$\Ar$,
\begin{multline}
	\nabla^2\Ar + k_0^2\Ar -\nabla\divrAr \\- j\omega\epsilon_0\mu_0\,\gradrphir =0,\quad\left(\vect{r}\in\mathcal{V}_0\right),\label{eq:pdeAout}
\end{multline}
where~${k_0 = \omega\sqrt{\epsilon_0\mu_0}}$ is the wave number in~$\mathcal{V}_0$.
Similarly, for~$\vect{r}\in\mathcal{V}$, we have
\begin{multline}
	\nabla^2\Ar + k^2\Ar -\nabla\divrAr \\- \left(j\omega\epsilon + \sigma\right)\mu\,\gradrphir = 0,\quad\left(\vect{r}\in\mathcal{V}\right),\label{eq:pdeAin}
\end{multline}
where~${k=\sqrt{-j\omega\mu\,(j\omega\epsilon + \sigma)}}$ is the wave number in~$\mathcal{V}$.
As in existing works~\cite{PIE04,PIE01,PIE02}, for~${\vect{r}\in\mathcal{V}_0}$, we use the usual Lorenz gauge
\begin{align}
	\divrAr = -j\omega\epsilon_0\mu_0\,\phir,\quad\left(\vect{r}\in\mathcal{V}_0\right),\label{eq:lorenzout}
\end{align}
in~\eqref{eq:pdeAout} to obtain the Helmholtz equation
\begin{align}
	\nabla^2\Ar + k_0^2\,\Ar = 0,\quad\left(\vect{r}\in\mathcal{V}_0\right).\label{eq:helmAout}
\end{align}
However, for~${\vect{r}\in\mathcal{V}}$, we propose a modified Lorenz gauge,
\begin{align}
	\divrAr = -\left(j\omega\epsilon + \sigma\right)\mu\,\phir[\mathrm{r}],\quad\left(\vect{r}\in\mathcal{V}\right),\label{eq:lorenzin}
\end{align}
where~$\phir[\mathrm{r}]$ is obtained by extracting the object's average surface potential from~$\phir$,
\begin{align}
	\phir[\mathrm{r}] &= \phir - \phi_{\mathrm{a}},\label{eq:phireddef}\\
	\phi_{\mathrm{a}} &= \frac{1}{\int_{\mathcal{S}}d\mathcal{S}}\,\int_{\mathcal{S}}d\mathcal{S}\,\phir.\label{eq:phiedef}
\end{align}
Using~\eqref{eq:lorenzin} in~\eqref{eq:pdeAin} and recalling that~$\phi_{\mathrm{a}}$ is a constant leads to the Helmholtz equation
\begin{align}
	\nabla^2\Ar + k^2\Ar = 0,\quad\left(\vect{r}\in\mathcal{V}\right).\label{eq:helmAin}
\end{align}

The modified gauge~\eqref{eq:lorenzin} is used for the following reasons:
in order to determine~$\phir$ uniquely within~$\mathcal{V}$, one must specify a boundary condition on~$\mathcal{S}$ and a reference point;
if the conventional Lorenz gauge
\begin{align}
	\divrAr = -\left(j\omega\epsilon + \sigma\right)\mu\,\phir,\quad\left(\vect{r}\in\mathcal{V}\right)\label{eq:lorenzinold}
\end{align}
is used, a change in reference for~$\phir$ will lead to a change in~$\divrArin{\mathcal{V}}$.
In other words,~$\divrArin{\mathcal{V}}$ cannot be determined uniquely until a reference is set for~$\phir$;
this is undesirable because it implies that~$\Arin{\mathcal{V}}$ may not be uniquely defined by~\eqref{eq:helmAin} even when the appropriate boundary conditions are provided, which implies a possible null space associated with the internal region, which can cause numerical issues.
Instead, when the constant average potential~$\phi_{\mathrm{a}}$ is extracted from~$\phir$, the remainder~$\phir[\mathrm{r}]$ becomes independent of the reference, as does~$\divrArin{\mathcal{V}}$ when the modified gauge~\eqref{eq:lorenzin} is used.
Equations~\eqref{eq:Adef} and~\eqref{eq:Atransform} reveal that~$\curlrArin{\mathcal{V}}$ is already independent of the reference for~$\phir$.
By ensuring that neither~$\divrArin{\mathcal{V}}$ nor~$\curlrArin{\mathcal{V}}$ depends on $\phi_{\mathrm{a}}$, the modified gauge~\eqref{eq:lorenzin} makes sure that~$\Arin{\mathcal{V}}$ is independent of the choice of reference for~$\phir$ and avoids the aforementioned null space.
The effectiveness of the modified gauge is demonstrated numerically in \secref{sec:results:int}: the condition number of the final system of equations with the modified gauge~\eqref{eq:lorenzin} is compared to the case when the conventional gauge~\eqref{eq:lorenzinold} is used; the latter leads to rank deficiency at low frequency.

The decomposition of~$\phir$ into~$\phi_{\mathrm{a}}$ and~$\phir[\mathrm{r}]$ and the use of the modified gauge~\eqref{eq:lorenzin} have an intuitive physical interpretation.
To compute the self capacitance of an object, its potential with respect to a reference must be known.
Instead, to compute its resistance or inductance, only the potential \emph{difference} across the object is involved.
The decomposition of~$\phir$ into~$\phi_{\mathrm{a}}$ and~$\phir[\mathrm{r}]$ maps to this physical interpretation:~$\phi_{\mathrm{a}}$ captures the choice of reference and is related to the capacitance of~$\mathcal{V}$ with respect to that reference, while~$\phir[\mathrm{r}]$ captures the spatial variation of~$\phir$ along~$\mathcal{S}$ independently of the reference, and so is related to the inductance and resistance of~$\mathcal{V}$.
Therefore, the modified gauge~\eqref{eq:lorenzin} simply implies that the inductive and resistive properties of~$\mathcal{V}$ are captured via~$\Arin{\mathcal{V}}$ and~$\phir[\mathrm{r}]$, while capacitive properties are modeled via~$\phi_{\mathrm{a}}$.

\subsection{Vector Potential Integral Equations}\label{sec:vpie}

As described in~\cite{PIE01}, the vector Green's second identity can be used along with~\eqref{eq:helmAin} to derive potential-based surface integral equations in terms of~$\Ar$ and its derivatives for both the internal and external regions.

\subsubsection{Internal region}\label{sec:vpie:int}
For the internal region, we have~\cite{PIE01}
 \begin{multline}
     \opL\bigl[\nhat'\times\curlrpArp\bigr] + \opK\bigl[\Atrp\bigr]
     + \Ar 
     \\- \opL\bigl[\divrpArp\,\nhat'\bigr]
     -\nabla\opL\bigl[\Anrp\bigr] = 0,\label{eq:vpieint}
 \end{multline}
where~$\vect{r}\in\mathcal{V}$,~$\vecprime{r}\in\mathcal{S}^-$, and primed (unprimed) coordinates denote source (test) points, respectively.
The integral operators in~\eqref{eq:vpieint} are defined as~\cite{ChewWAF}
 \begin{align}
 	\opL\bigl[\vect{a}\rp\bigr] &= \int_{\mathcal{S}}d\mathcal{S}'\,\Grrpk\,\vect{a}\rp,\label{opLdef}\\
 	\opK\bigl[\vect{a}\rp\bigr] &= \int_\mathcal{S}d\mathcal{S}'\,\nabla\Grrpk\times\vect{a}\rp,\label{opKdef}
 \end{align}
 where~$\Grrpk$ is the Green's function associated with the material in~$\mathcal{V}$,
 \begin{align}
 	\Grrpk = \frac{e^{-jk\abs{\vect{r}-\vecprime{r}}}}{4\pi\abs{\vect{r}-\vecprime{r}}}.\label{eq:grrpk}
 \end{align}
Two separate integral equations can be derived from~\eqref{eq:vpieint} by taking its tangential and normal components~\cite{PIE01}.
Applying the modified Lorenz gauge~\eqref{eq:lorenzin} and letting~$\vect{r}\to\mathcal{S}^-$, the tangential and normal components of~\eqref{eq:vpieint} become, respectively,
 \begin{multline}
	\nhat\times\opL\Bigl[\nhat'\times\curlrpArp\Bigr] + \nhat\times\opKpv\Bigl[\Atrp\Bigr]\\
	+ \frac{1}{2}\,\Atr
	+ \left(j\omega\epsilon + \sigma\right)\mu\,\nhat\times\opL\Bigl[\phirp[\mathrm{r}]\,\nhat'\Bigr]\\
	-\nhat\times\nabla\opL\Bigl[\Anrp\Bigr] = 0,\label{eq:tvpieint}
\end{multline}
and
 \begin{multline}
	\nhat\cdot\opL\Bigl[\nhat'\times\curlrpArp\Bigr] + \nhat\cdot\opKpv\Bigl[\Atrp\Bigr]\\
	+ \left(j\omega\epsilon + \sigma\right)\mu\,\nhat\cdot\opL\Bigl[\phirp[\mathrm{r}]\,\nhat'\Bigr]\\
	- \opMdpv\Bigl[\Anrp\Bigr] + \frac{1}{2}\,\Anr = 0,\label{eq:nvpieint}
\end{multline}
where
 \begin{align}
	\opMdpv\Bigl[a\rp\Bigr] = \pvint_{\mathcal{S}}d\mathcal{S}'\,\nhat\cdot\gradrGrrpk\,\vect{a}\rp.\label{opMdpvdef}
\end{align}
In~\eqref{eq:tvpieint}~\eqref{eq:nvpieint}, and~\eqref{opMdpvdef}, the symbol~$\pvint$ and dashes through the operators indicate that the associated integrals are computed in a principal value sense~\cite{ChewWAF}.

\subsubsection{External region}\label{sec:vpie:ext}
\begin{figure}[t]
	\centering
	\includegraphics[width=0.8\linewidth]{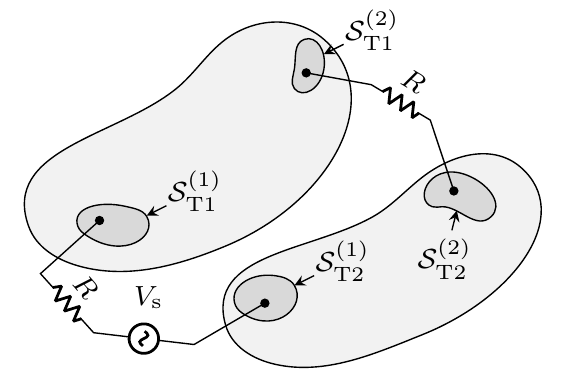}
	\caption{Two-port network considered in \secref{sec:vpie:ext}.}\label{fig:portgeom}
\end{figure}

The coupling between the electromagnetic and circuit problems can be modeled through the external problem.
We assume that the structure to be simulated contains one or more electrically small lumped ports, where each port consists of two terminals connected to a Th\'evenin equivalent circuit~\cite{gope}, with resistance~$R$ and voltage source $V_{\mathrm{s}}$.
The approach described here is applicable for any number of ports and objects, but we consider the setup in \figref{fig:portgeom} for simplicity, which shows a two-port network with two objects and a voltage source at port~$1$.

A vector potential integral equation analogous to~\eqref{eq:vpieint} can be derived for the external region,
 \begin{multline}
	\opL[0]\bigl[\nhat'\times\curlrpArp\bigr] + \opK[0]\bigl[\Atrp\bigr]
	- \Ar 
	\\- \opL[0]\bigl[\divrpArp\,\nhat'\bigr]
	-\nabla\opL[0]\bigl[\Anrp\bigr] = 0.\label{eq:vpieext}
\end{multline}
Taking the tangential component of~\eqref{eq:vpieext} and applying the Lorenz gauge~\eqref{eq:lorenzout}, the boundary condition~\eqref{eq:bcAnr}, and the decomposition~$\eqref{eq:phireddef}$,
\begin{multline}
	\nhat\times\opL[0]\Bigl[\nhat'\times\curlrpArp\Bigr] \\+ \nhat\times\opKpv[0]\Bigl[\Atrp\Bigr]
	- \frac{1}{2}\,\Atr
	\\+ j\omega\epsilon_0\mu_0\,\nhat\times\opL[0]\Bigl[(\phirp[\mathrm{r}] + \phi_\mathrm{a})\,\nhat'\Bigr] = 0,\label{eq:tvpieext}
\end{multline}
where~$\vect{r},\vecprime{r}\in\mathcal{S}^+$, and the subscript~``$0$'' on the operators indicates that the Green's function associated with~$\mathcal{V}_0$ is used.
Unlike for the internal problem, we next take the divergence of~\eqref{eq:vpieext}~\cite{PIE03},
\begin{multline}
	\opL[0]\Bigl[\nabla'\cdot\left(\nhat'\times\curlrpArp\right)\Bigr]
	\\+ j\omega\epsilon_0\mu_0\,\opMpv[0]\Bigl[(\phirp[\mathrm{r}] + \phi_{\mathrm{a}})\Bigr] + \frac{1}{2}\,j\omega\epsilon_0\mu_0\left(\phirp[\mathrm{r}] + \phi_{\mathrm{a}}\right) 
	\\ + \opL[0]\Bigl[J_\mathrm{T}\rp\Bigr] = 0,\label{eq:dvpieext}
\end{multline}
rather than its normal component.
In~\eqref{eq:dvpieext}, equations~\eqref{eq:lorenzout},~\eqref{eq:bcAnr}, and~$\eqref{eq:phireddef}$ were all used, and
\begin{align}
	\opMpv\Bigl[a\rp\Bigr] = \pvint_{\mathcal{S}}d\mathcal{S}'\,\nhat'\cdot\gradrGrrpk\,\vect{a}\rp.\label{opMpvdef}
\end{align}
The divergence of the vector potential integral equation~\eqref{eq:dvpieext} is related to the 2D continuity equation on~$\mathcal{S}^+$ in weak form~\cite{PIE03}.
This allows us to introduce in~\eqref{eq:dvpieext} the terminal volume current density~$J_\mathrm{T}\r$ associated with the attached circuit, where~$J_\mathrm{T}\r$ is normal to~$\mathcal{S}^+$ and is non-zero only for~$\vect{r}\in\mathcal{S}_{\mathrm{T}}^+$.
Therefore, the external circuit can be incorporated in a straightforward manner by introducing an additional set of equations to relate~$J_\mathrm{T}\r$ in~\eqref{eq:dvpieext} to the connected circuit~\cite{gope}.
The additional equations for~$J_\mathrm{T}\r$ are obtained by writing the KVL for each terminal of port~$i$,
\begin{align}
	\phirin{\mathcal{S}_{\mathrm{T}1}^{(i)}} - \phirin{\mathcal{S}_{\mathrm{T}2}^{(i)}} + J_{\mathrm{T}1}^{(i)}A_{\mathrm{T}1}^{(i)}R &= V_{\mathrm{s}}^{(i)},\label{eq:kvlT1}\\
	\phirin{\mathcal{S}_{\mathrm{T}1}^{(i)}} - \phirin{\mathcal{S}_{\mathrm{T}2}^{(i)}} - J_{\mathrm{T}2}^{(i)}A_{\mathrm{T}2}^{(i)}R &= V_{\mathrm{s}}^{(i)},\label{eq:kvlT2}
\end{align}
where superscript~$(i)$ indicates that the associated quantity is related to port~$(i)$.
In~\eqref{eq:kvlT1} and~\eqref{eq:kvlT2},~$\mathcal{S}_{\mathrm{T}t}^{(i)}$ denotes terminal~$t$ of port~$i$, and~$A_{\mathrm{T}t}^{(i)}$ is the area of~$\mathcal{S}_{\mathrm{T}t}^{(i)}$.
In~\eqref{eq:kvlT1} and~\eqref{eq:kvlT2},~$J_{\mathrm{T}t}^{(i)}$ is the normal component of the volume current density flowing into terminal~$t$ of port~$i$.

\subsection{Charge Neutrality Enforcement}\label{sec:charge}

At very low frequencies, and particularly at DC, charge neutrality must be enforced for numerical stability.
A similar requirement exists for several field-based formulations~\cite{aefie2,eaefie01,AWPLSLIM,JMMCTSLDM}.
However, unlike in those methods, the electric surface charge density is not available as an unknown in the proposed approach.
Therefore, we enforce charge neutrality indirectly through~$\phi_{\mathrm{a}}$.
For a single conductive object at DC (\figref{fig:geom}), the capacitance~$C$ with respect to infinity can be defined via
\begin{align}
	C\phi_{\mathrm{a}} = Q,\label{eq:cap}
\end{align}
where~$Q$ is the total charge on the object.
For charge neutrality, we require~${Q=0}$, which implies~${\phi_{\mathrm{a}}=0}$.
However, we are concerned with multiple objects connected to each other through ports, which requires a more general treatment.
If~$N_{\mathrm{c}}$ objects are connected through ports (e.g., \figref{fig:portgeom}, where~${N_{\mathrm{c}}=2}$), we can define a capacitance~$C_{\mathrm{c}}$ associated with the entire set with respect to infinity as
\begin{align}
	C_{\mathrm{c}}\phi_{\mathrm{ac}} = Q_{\mathrm{c}},\label{eq:capset}
\end{align}
where
\begin{align}
	Q_{\mathrm{c}} &= \sum_i^{N_{\mathrm{c}}}Q_i,\label{eq:Qset}\\
	\phi_{\mathrm{ac}} &= \frac{1}{N_{\mathrm{c}}}\sum_i^{N_{\mathrm{c}}}\phi_{\mathrm{a}i},\label{eq:phieset}
\end{align}
and~$\phi_{\mathrm{a}i}$ is the average surface potential of object~$i$.
From~\eqref{eq:capset}, we see that~${Q_{\mathrm{c}} = 0}$ can be enforced by requiring
\begin{align}
	\phi_{\mathrm{ac}} = 0.\label{eq:neutr}
\end{align}
Equation~\eqref{eq:neutr} is necessary only at low frequencies, so we enforce it when the structure's electrical diameter is less than~$0.1\,\lambda_0$, where~${\lambda_0=2\pi/k_0}$ is the wavelength in free space.
A more general charge conservation condition can also be applied when a non-zero total charge~$Q_{\mathrm{c}}$ is to be specified, by taking~$\nhat\cdot\gradrphirin{\mathcal{S}^+}$ as an additional unknown and solving the scalar potential integral equation in addition to the vector-potential based equations described in \secref{sec:vpie}~\cite{PIE01,AWPLVPIE}.
If the neutrality condition~\eqref{eq:neutr} is not enforced, the formulation becomes poorly conditioned at low frequency, as shown numerically in \secref{sec:results:int}.

\subsection{Discretization}\label{sec:discr}

For convenience, we define~${\mathcal{S} = \bigcup_{i}^{N_{\mathrm{obj}}}\mathcal{S}_{i}}$, where~$\mathcal{S}_{i}$ is the surface of object~$i$.  ($N_{\mathrm{obj}}=2$ for the case of~\figref{fig:portgeom}).

\subsubsection{Choice of basis and testing functions}\label{sec:discr:basis}
A triangular mesh is generated for~$\mathcal{S}$, and we adopt the discretization scheme proposed in~\cite{AWPLVPIE} for the vector potential integral equations~\eqref{eq:tvpieint},~\eqref{eq:nvpieint},~\eqref{eq:tvpieext}, and~\eqref{eq:dvpieext}.
Quantity~${\nhat\times\curlrAr}$ is expanded with Rao-Wilton-Glisson (RWG) functions~\cite{RWG}~$\vect{f}_n\r$ normalized by edge length.
Instead,~$\Atr$ is expanded with Buffa-Christiansen functions~\cite{BCorig}~$\vect{g}_n\r$, which are defined on a barycentric refinement of the mesh.
This discretization scheme ensures that the mutual orthogonality of~$\nhat\times\curlrAr$ and~$\Atr$ is respected, eventually leading to well tested operators~\cite{eaefie01}.
Scalar quantities~$\phir$ and~$\Anr$ are expanded with unit-amplitude pulse functions ~$h_n\r$.
Using nodal linear functions to expand~$\phir$ may lead to better matrix conditioning and smaller errors than with pulse functions~\cite{PIE_TD_DIEL}, but the proposed discretization scheme still provides excellent accuracy for complex structures, as verified in \secref{sec:results}.
Finally, the normal component of the terminal current density~$J_{\mathrm{T}}\r$ is expanded with normalized pulse functions~$h_n\r/A_n$, where~$A_n$ is the area of the~$n\text{th}$ triangle.
We assume that each terminal is associated with a single mesh triangle, so that there are as many unknown coefficients associated with~$J_{\mathrm{T}}\r$  as the number of terminals.
This assumption can be relaxed easily by introducing an additional equation to enforce a constant scalar potential on all triangles associated with a terminal~\cite{gope}.

\subsubsection{Scalar potential remainder term}\label{sec:discr:phi}
In order to enforce~\eqref{eq:neutr}, and considering~\eqref{eq:phireddef}, we take~$\phir[\mathrm{r}]$ and~$\phi_{\mathrm{a}}$ as separate unknowns.
From~\eqref{eq:phireddef}, it is apparent that~$\phirin[\mathrm{r}]{\mathcal{S}_i}$ has a zero average value over~$\mathcal{S}_i$ because it represents the remainder after extracting the average surface potential of that object~$\phi_{\mathrm{a}}^{(i)}$.
The choice of basis function for~$\phirin[\mathrm{r}]{\mathcal{S}_i}$ must preserve this zero-mean property.
In the following, we assume that the mesh for object~$i$ contains~$N_{\mathrm{tri}}^{(i)}$ triangles, and column vector~${\Phimat[\mathrm{r}]^{(i)}}$ contains unknown coefficients associated with~$\phirin[\mathrm{r}]{\mathcal{S}_i}$.
To preserve the zero-mean property of~$\phirin[\mathrm{r}]{\mathcal{S}_i}$, vector~${\Phimat[\mathrm{r}]^{(i)}}$ should belong to the subspace~$\mathbb{C}^{(N_{\mathrm{tri}}^{(i)}-1)}$.
Accordingly, we seek a basis~$\matr{D}_{\mathrm{r}}^{(i)}$ of dimension~${N_{\mathrm{tri}}^{(i)}-1}$ to expand~$\phirin[\mathrm{r}]{\mathcal{S}_i}$.
Then, the discrete counterpart of~\eqref{eq:phireddef} can be written for each object~$i$ as
\begin{align}
	\matr{D}_{\mathrm{r}}^{(i)}\Phimat[\mathrm{r}]^{(i)} = \Phimat^{(i)} - \1^{(i)}\phi_\mathrm{a}^{(i)},\label{eqd:phirt}
\end{align}
where~${\Phimat[\mathrm{r}]^{(i)}\in\mathbb{C}^{(N_{\mathrm{tri}}^{(i)}-1)}}$ and~${\Phimat^{(i)}\in\mathbb{C}^{N_{\mathrm{tri}}^{(i)}}}$.
In~\eqref{eqd:phirt},~$\Phimat^{(i)}$ contains the unknown coefficients associated with~$\phirin{\mathcal{S}_i}$, and column vector~${\1^{(i)}\in\mathbb{R}^{N_{\mathrm{tri}}^{(i)}}}$ contains all ones.
Matrix~${\matr{D}_{\mathrm{r}}^{(i)}\in\mathbb{R}^{N_{\mathrm{tri}}^{(i)}\times (N_{\mathrm{tri}}^{(i)}-1)}}$ is sparse and will be defined below.
To check that the zero-mean property of~$\phirin[\mathrm{r}]{\mathcal{S}_i}$ is preserved in the discrete domain, we can left-multiply~\eqref{eqd:phirt} by~$(\1^{(i)})^T$,
\begin{align}
	(\1^{(i)})^T\matr{D}_{\mathrm{r}}^{(i)}\Phimat[\mathrm{r}]^{(i)} = (\1^{(i)})^T\Phimat^{(i)} - (\1^{(i)})^T\1^{(i)}\phi_\mathrm{a}^{(i)},\label{eqd:1Tphirt}
\end{align}
which is the discrete equivalent to an integral over~$\mathcal{S}_i$.
The first term on the right-hand side of~\eqref{eqd:1Tphirt} can be written as
\begin{align}
	(\1^{(i)})^T\Phimat^{(i)} = N_{\mathrm{tri}}^{(i)}\,\phi_\mathrm{a}^{(i)}\label{eqd:1Tphirt1}
\end{align}
because of the definition of~$\phi_\mathrm{a}^{(i)}$, and the second term on the right-hand side of~\eqref{eqd:1Tphirt} is
\begin{align}
	(\1^{(i)})^T\1^{(i)}\phi_\mathrm{a}^{(i)} = N_{\mathrm{tri}}^{(i)}\,\phi_\mathrm{a}^{(i)}.\label{eqd:1Tphirt2}
\end{align}
Using~\eqref{eqd:1Tphirt1} and~\eqref{eqd:1Tphirt2} in~\eqref{eqd:1Tphirt} immediately reveals that
\begin{align}
	(\1^{(i)})^T\matr{D}_{\mathrm{r}}^{(i)}\Phimat[\mathrm{r}]^{(i)} = 0,
\end{align}
as desired.

There are several possible choices for~$\matr{D}_{\mathrm{r}}^{(i)}$ which satisfy~\eqref{eqd:phirt}.
Here, we choose
\begin{align}
	\matr{D}_{\mathrm{r}}^{(i)} \triangleq \begin{bmatrix} \matr{I}_{\mathrm{r}} \\ -\left(\1^{(i)}\right)^T \end{bmatrix},\label{eqd:Drdef}
\end{align}
where~${\matr{I}_{\mathrm{r}}\in\mathbb{R}^{(N_{\mathrm{tri}}^{(i)}-1) \times (N_{\mathrm{tri}}^{(i)}-1)}}$ is the identity matrix.
Physically, this choice implies that the entries of~$\Phimat[\mathrm{r}]^{(i)}$ represent potentials relative to the average surface potential on~$\mathcal{S}_i$.
Finally, the vectors of scalar potential unknowns associated with each object are concatenated together by defining
\begin{align}
	\Phimat[\mathrm{r}] =
	\begin{bmatrix}
		\Phimat[\mathrm{r}]^{(1)} \\ \Phimat[\mathrm{r}]^{(2)}
	\end{bmatrix},\quad
	\Phimat[\mathrm{a}] =
	\begin{bmatrix}
		\phi_{\mathrm{a}}^{(1)} \\ \phi_{\mathrm{a}}^{(2)}
	\end{bmatrix},\label{eqd:phiremat}
\end{align}
so that
\begin{align}
	\Phimat &= \matr{D}_{\mathrm{r}}\Phimat[\mathrm{r}] + \1\Phimat[\mathrm{a}],\label{eqd:phimat}
\end{align}
where
\begin{align}
	\matr{D}_{\mathrm{r}} &=
	\begin{bmatrix}
		\matr{D}_{\mathrm{r}}^{(1)} & \matr{0} \\
		\matr{0} & \matr{D}_{\mathrm{r}}^{(2)}
	\end{bmatrix},\label{eqd:Dr}\\
	\1 &=
	\begin{bmatrix}
		\1^{(1)} & \matr{0} \\
		\matr{0} & \1^{(2)}
	\end{bmatrix}.\label{eqd:1}
\end{align}

\subsection{Final System of Equations}\label{sec:discr:sys}
The vector equations~\eqref{eq:tvpieint} and~\eqref{eq:tvpieext} are tested with~$\nhat\times\text{RWG}$ functions, while the scalar equations~\eqref{eq:nvpieint} and~\eqref{eq:dvpieext} are tested with~$h_n\r/A_n$.
To assemble the final system of equations, the discrete versions of~\eqref{eq:tvpieext},~\eqref{eq:tvpieint},~\eqref{eq:dvpieext},~\eqref{eq:nvpieint}, and the KVL equations~\eqref{eq:kvlT1} and~\eqref{eq:kvlT2}, are concatenated.
The resulting system of equations~\eqref{eqd:sys2} is at the top of the following page.
\begin{figure*}[ht!]
	\normalsize
	\setcounter{equation}{\getrefnumber{eqd:1}}
	\begin{align}
		{\renewcommand*{\arraystretch}{1.25}
			\begin{bmatrix}
				\frac{1}{\xi}\Lttmat[0] & \Kttmat[0] & \frac{jk_0}{\xi}\,\Ltnmat[0]\matr{D}_{\mathrm{r}} & \frac{jk_0}{\xi}\,\Ltnmat[0]\1 & \matr{0} & \matr{0} \\
				\frac{\mu}{\xi\mu_0}\Lttmat & \Kttmat & \frac{c_0\gamma}{\xi}\,\Ltnmat\matr{D}_{\mathrm{r}} & \matr{0} & \frac{1}{\xi}\matr{D}^T\Lssmat & \matr{0} \\
				\xi\,\matr{F}\Lssmat[0]\matr{D} & \matr{0} & \xi\,jk_0\,\matr{F}\Msnmat[0]\matr{D}_{\mathrm{r}} & \xi\,jk_0\,\matr{F}\Msnmat[0]\1 & \matr{0} & \xi\,\matr{F}\Lssmat[0]\matr{D}_{\mathrm{T}} \\
				\frac{\mu}{\mu_0}\Lntmat & \xi\,\Kntpvmat & c_0\gamma\,\Lnnmat\matr{D}_{\mathrm{r}} & \matr{0} & \Msnmat & \matr{0} \\
				\matr{0} & \matr{0} & \matr{0} & \matr{S} & \matr{0} & \matr{0} \\			%
				\matr{0} & \matr{0} & \frac{1}{c_0}\,\matr{P}\matr{D}_{\mathrm{r}} & \frac{1}{c_0}\,\matr{P}\1 & \matr{0} & \frac{1}{\eta_0}\,\matr{R}
		\end{bmatrix}}
		{\renewcommand*{\arraystretch}{1.25}
			\begin{bmatrix}
				\uamat[0] \\ \ubmat/\xi \\ \Phimat[\mathrm{r}]/c_0 \\ \Phimat[\mathrm{a}]/c_0 \\ \udmat \\ \mu_0\,\Jmat[\mathrm{T}]
		\end{bmatrix}}
		=
		{\renewcommand*{\arraystretch}{1.25}
			\begin{bmatrix}
				\matr{0} \\ \matr{0} \\ \matr{0} \\ \matr{0} \\ \matr{0} \\ \matr{V}_{\mathrm{s}}/c_0
		\end{bmatrix}}.%
		\label{eqd:sys2}
	\end{align}
	\hrulefill
\end{figure*}
In~\eqref{eqd:sys2}, we have introduced the following symbols for simplicity,
\begin{align}
	\Kttmat[0] &= \Kttpvmat[0] - \frac{1}{2}\Irtmat,\label{eqd:Kttmat0}\\
	\Kttmat &= \Kttpvmat + \frac{1}{2}\Irtmat,\label{eqd:Kttmat}\\
	\Msnmat[0] &= \Msnpvmat[0] + \frac{1}{2}\Issmat,\label{eqd:Msnmat0}\\
	\Msnmat &= \bigl(\Mnspvmat\bigr)^T - \frac{1}{2}\Issmat,\label{eqd:Msnmat}
\end{align}
where in~\eqref{eqd:sys2}--\eqref{eqd:Msnmat},~$\Lmat$,~$\Kpvmat$ and~$\Mpvmat$ are the discretized~$\opL$,~$\opKpv$ and~$\opMpv$ operators, respectively.
The superscript labels~$(mn)$ on each matrix operator represent the testing and basis functions involved, respectively.
Operators~$\Irtmat$ and~$\Issmat$ are Gram matrices linking the associated basis and testing function spaces.
The matrix operators associated with the internal problem, which appear in the second and fourth equations in~\eqref{eqd:sys2}, have a block diagonal structure where the number of blocks equals~$N_{\mathrm{obj}}$, because they are local to each object.
Term~$\matr{D}$ is a sparse incidence matrix linking mesh edges and triangles, whose definition can be found in~\cite{aefie2}, and~$\matr{D}_{\mathrm{T}}$ is a sparse incidence matrix which selects triangles associated with terminals~\cite{gope}.
Matrix~$\matr{P}$ contains negative and positive ones to compute potential differences between the terminals of each port, and~$\matr{R}$ contains the resistance associated with the Th\'evenin equivalent circuit attached to each port~\cite{gope}.

The fifth equation in~\eqref{eqd:sys2} is the discrete counterpart of the charge neutrality condition~\eqref{eq:neutr}, where~$\matr{S}$ contains ones and adds the average potential of both objects.
This equation is only included when the structure's electrical diameter is smaller than~$0.1\,\lambda_0$ as mentioned in \secref{sec:charge}.
In that case, the system is no longer square, and a corresponding number of equations is deleted from the third row in~\eqref{eqd:sys2}, which is the discrete version of~\eqref{eq:dvpieext}.
The deletion is accomplished with the sparse matrix~$\matr{F}$, which removes one equation per set of connected objects. For example, the two objects in \figref{fig:portgeom} are connected by ports and therefore part of a set; only one equation is deleted for the pair~\cite{aefie2}.
For isolated objects not connected to any port, one equation is deleted per object.
When the electrical diameter of the structure is larger than~$0.1\,\lambda_0$, the fifth equation and~$\matr{F}$ are excluded from the system.
Column vectors~$\uamat$,~$\ubmat$,~$\udmat$, and~$\Jmat[\mathrm{T}]$ contain the unknown coefficients associated with~${\nhat\times\curlrAr}$,~${\Atr}$,~${\Anr}$, and~${\Jr[\mathrm{T}]}$, respectively.
In order to obtain a stable condition number even at low frequencies, the equations and unknowns are scaled strategically, as suggested in~\cite{AWPLVPIE}.
Quantity~$\xi$ is the average mesh edge length,~${\gamma_0 = j\omega\epsilon_0\mu_0}$, and~${\gamma = \left(j\omega\epsilon + \sigma\right)\mu}$.
If different objects are composed of different materials, the appropriate value of~$\gamma$ associated with each material should be used.
With this scaling scheme, each block in the system matrix in~\eqref{eqd:sys2} is dimensionless, while each vector in the list of unknowns has units of the magnetic vector potential,~$\text{V}\cdot \text{s}/\text{m}$.
The condition number of the system matrix in~\eqref{eqd:sys2} is reported from DC to high frequencies for some of the numerical examples in \secref{sec:results}, and remains stable down to DC, unlike many field-based formulations~\cite{lfbreakdown}.
In particular, the results and analysis of condition numbers in \secref{sec:results} show that the proposed formulation is accurate and has full rank down to DC, and is amenable to the use of an iterative solver with an appropriate preconditioner.
We also demonstrate in \secref{sec:results:int} that the modified gauge~\eqref{eq:lorenzin} and charge neutrality condition~\eqref{eq:neutr} are crucial; without them, the formulation becomes rank deficient at low frequency.


\subsection{Implementation Considerations}\label{sec:discr:impl}

In~\eqref{eqd:sys2}, some matrix blocks such as~$\Lntmat$ and~$\Kntpvmat$ involve basis functions tangential to~$\mathcal{S}$ and testing functions normal to~$\mathcal{S}$, or vice versa.
In these cases, the testing functions may not be in the correct range space for the associated integral operator~\cite{KleinmanErrCond}, so the corresponding matrix block may be poorly conditioned.
An improved discretization scheme is beyond the scope of this work; instead, we ensure that the entries of these poorly tested matrices are computed with high accuracy, particularly for the matrix blocks associated with the internal region, to mitigate the effect of poor conditioning on the overall accuracy.
To this end, we employ the polar-coordinate integration technique proposed in~\cite{gibc} for all discrete integral operators in~\eqref{eqd:sys2} which are associated with the internal region, even at low frequencies.

For the polar-coordinate integration scheme applied to internal region operators, we use Gaussian quadrature of order~$14$ for the integration over edges of the source triangle~\cite{gibc}.
For the test triangle, we use a cubature rule which depends on the distance between the test and source triangles.
For test triangles which share at least one vertex with the source triangle, or are within~$5\,\xi$ of the source triangle, we use a cubature rule of order~$25$.
For all other test triangles we use an order of~$13$.
The numerical results are fairly insensitive to the integration order for matrix operators associated with the external region, and standard quadrature rules and singularity extraction techniques can be used~\cite{gibson}.
The stronger requirement on integration accuracy of the proposed method compared to field-based methods is made worthwhile by DC-to-high-frequency modeling capability, unlike conventional BEM formulations.
An improved discretization scheme would be desirable to relax these requirements on integration accuracy, and will be considered in future work.
The numerical examples presented in \secref{sec:results} show that accurate results are obtained despite the presence of some poorly tested matrix blocks.

\section{Results}\label{sec:results}

Numerical examples drawn from chip- and package-level applications are presented here.
The scattering~($S$) parameters computed via the proposed method~\eqref{eqd:sys2} are compared against those obtained via the commercial finite element solver~Ansys HFSS, and two state-of-the-art field-based BEM formulations which are stable at low frequency by design: the augmented generalized impedance boundary condition~(AGIBC)~\cite{gibc,agibc} and the enhanced augmented electric field integral equation~(eAEFIE)~\cite{eaefie01,eaefie02}.
Though these field-based BEM formulations surpass many others in their ability to model conductors at low frequency, we show that they too eventually become inaccurate at very low frequencies, while the proposed method does not.
In all methods, a direct solver based on~LU factorization~\cite{trefethen} is used to solve the final system of equations.

\subsection{Interconnects With a Trapezoidal Cross Section}\label{sec:results:int}

\begin{figure}[t]
	\centering
	\includegraphics[width=.75\linewidth]{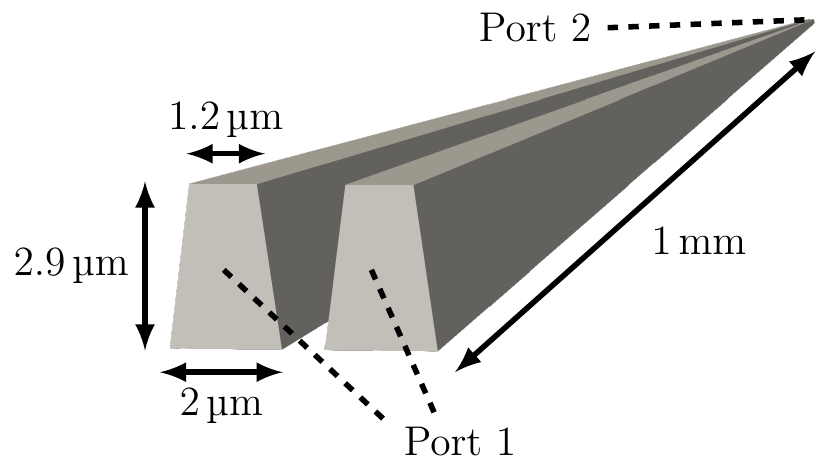}
	\caption{Geometry and port definitions for the interconnect in \secref{sec:results:int}.}\label{fig:int:geom}
\end{figure}

\begin{figure}[t]
	\centering
	\includegraphics[width=\linewidth]{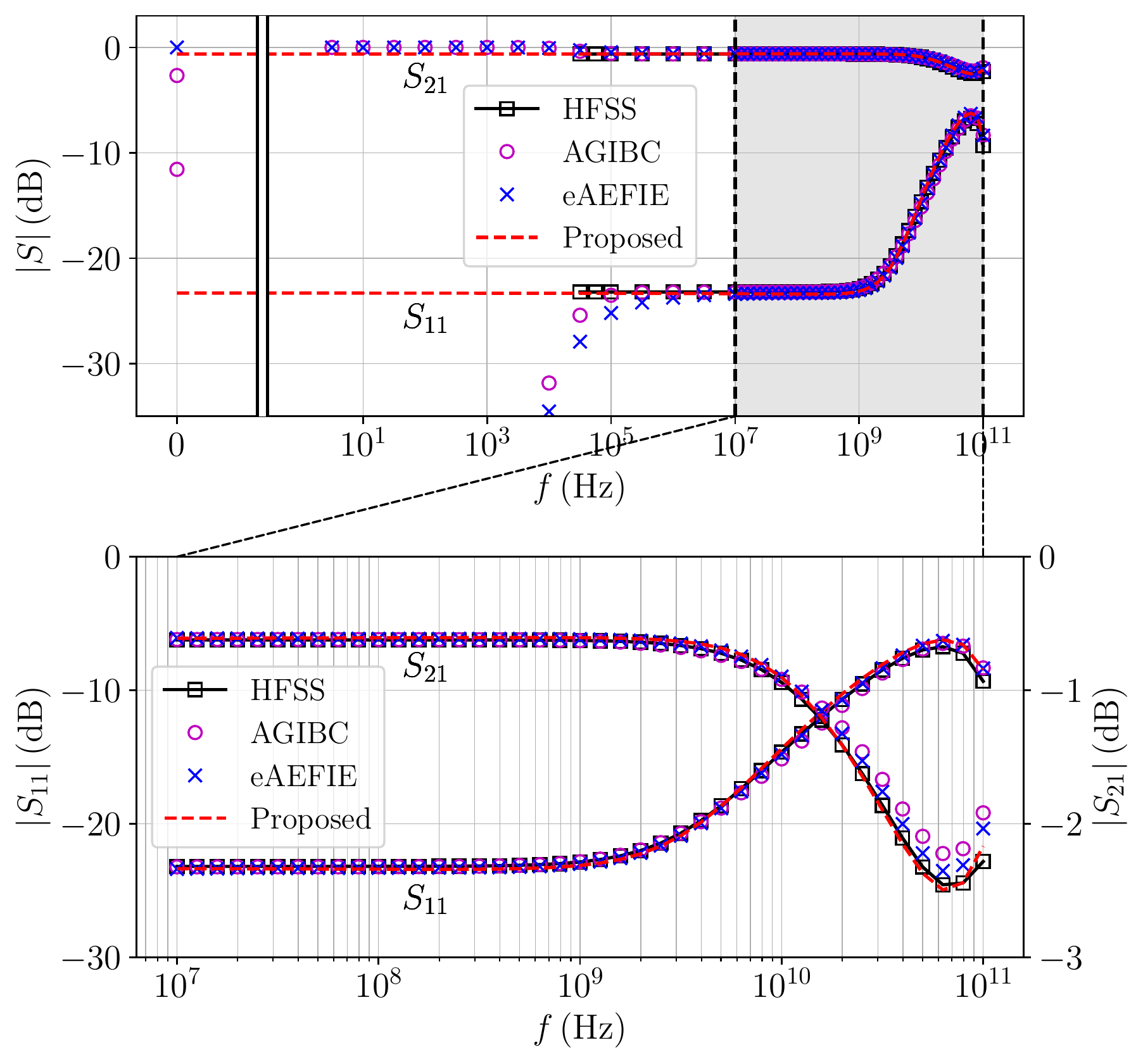}
	\caption{Scattering parameters (magnitude) for the interconnect in \secref{sec:results:int}.}\label{fig:int:Smag}
\end{figure}

\begin{figure}[t]
	\centering
	\includegraphics[width=\linewidth]{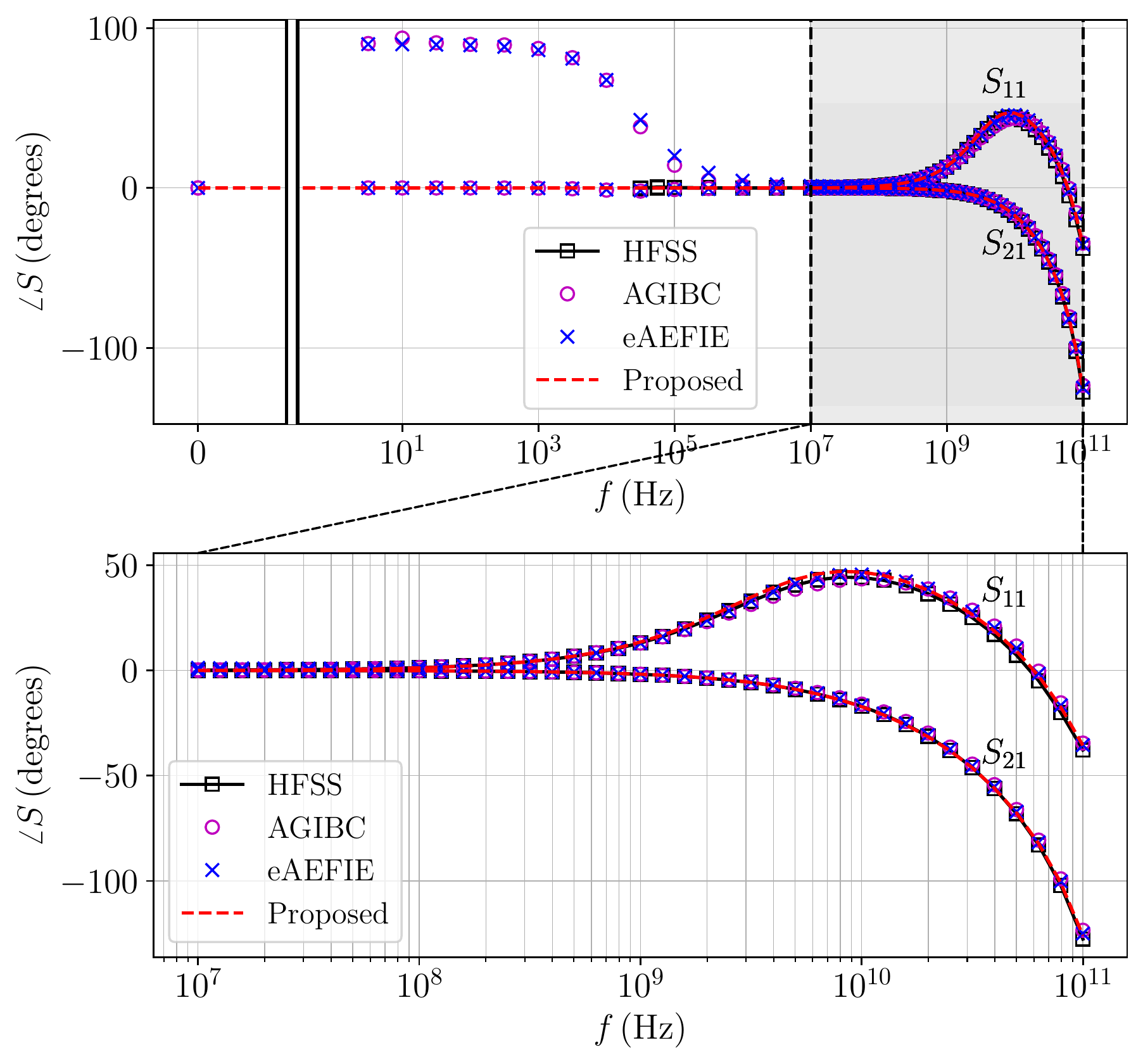}
	\caption{Scattering parameters (phase) for the interconnect in \secref{sec:results:int}.}\label{fig:int:Sang}
\end{figure}

We first consider a pair of copper conductors~$1\,$mm long with a trapezoidal cross section~(\figref{fig:int:geom}), meshed with~$2{,}008$ triangles.
The structure is excited with a differential signal via two ports at either end, as shown in~\figref{fig:int:geom}.
\figref{fig:int:Smag} and \figref{fig:int:Sang} show, respectively, the magnitude and phase of the~$S$ parameters for each of the methods considered.
In both \figref{fig:int:Smag} and \figref{fig:int:Sang}, the top panel shows the entire frequency range considered, from DC to~$100\,$GHz, while the bottom panel focuses on the high-frequency regime to better resolve the sharper variations in the~$S$ parameters.
As frequency decreases, the~$S$ parameter magnitude is expected to remain constant, while the phase should approach~$0^{\circ}$.
The proposed method is the only one which remains accurate in terms of both magnitude and phase down to DC.
In particular, HFSS aborts with a solver error below~${\sim}10\,$kHz, while the field-based methods become inaccurate below~${\sim}100\,$kHz.
Moreover, since the proposed full-wave method does not employ any quasistatic approximations, it also remains accurate at very high frequencies up to~$100\,$GHz.
\figref{fig:int:Smag} and \figref{fig:int:Sang} thus demonstrate the excellent broadband capabilities of the proposed potential-based formulation.

\begin{figure}[t]
	\centering
	\includegraphics[width=\linewidth]{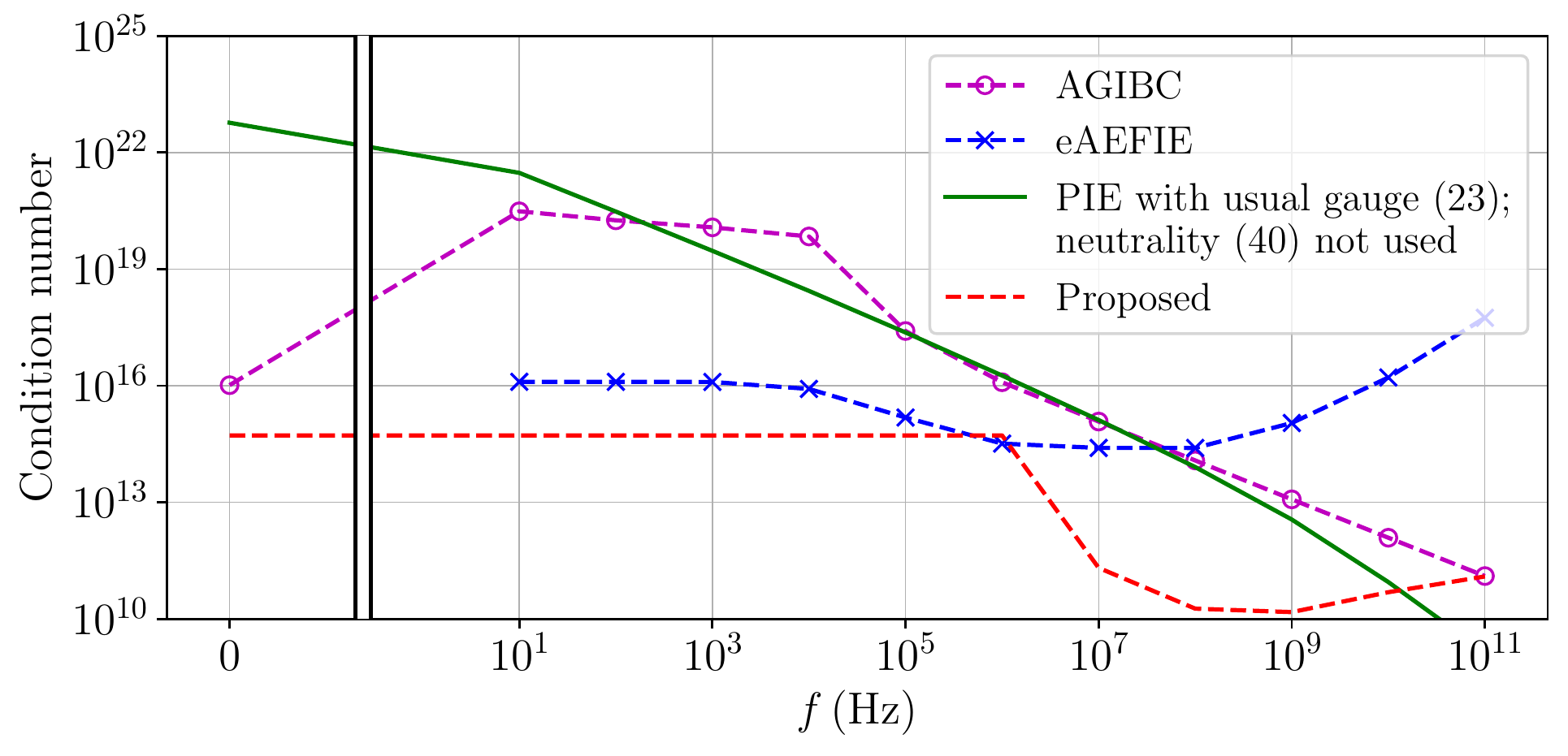}
	\caption{Condition number of the system matrix in~\eqref{eqd:sys2} for the interconnect in \secref{sec:results:int}, compared to that of field-based formulations and the case when the proposed modified gauge~\eqref{eq:lorenzin} and neutrality condition~\eqref{eq:neutr} are not used.}\label{fig:int:cond}
\end{figure}

We next studied the condition number of the system matrix in~\eqref{eqd:sys2}, as shown in \figref{fig:int:cond}.
To demonstrate the need and importance of the modified gauge~\eqref{eq:lorenzin} and the charge neutrality condition~\eqref{eq:neutr}, we compared the proposed approach to the case when the conventional gauge~\eqref{eq:lorenzinold} is used in the internal problem, and the charge neutrality condition~\eqref{eq:neutr} is not applied.
\figref{fig:int:cond} shows that the condition number of the proposed system matrix in~\eqref{eqd:sys2} remains stable over the entire frequency range down to exactly DC.
Instead, without~\eqref{eq:lorenzin} and~\eqref{eq:neutr}, the condition number increases dramatically as frequency decreases and the system becomes rank deficient.
Although the condition number associated with the proposed method is large~(${\sim}10^{14}$), a key point is that it remains stable and can likely be reduced further with preconditioning techniques.
For perspective, the condition numbers associated with the AGIBC and eAEFIE are also shown in~\figref{fig:int:cond}.
Even at moderately high frequencies, the condition number associated with the field-based formulations is large~(${\sim}10^{13}$--$10^{14}$) and a preconditioner is required in order to use an iterative solver~\cite{aefie2,eaefie01}.

\subsection{Differential Pair With a Microvia Array}\label{sec:results:microvia}

\begin{figure}[t]
	\centering
	\includegraphics[width=1.0\linewidth]{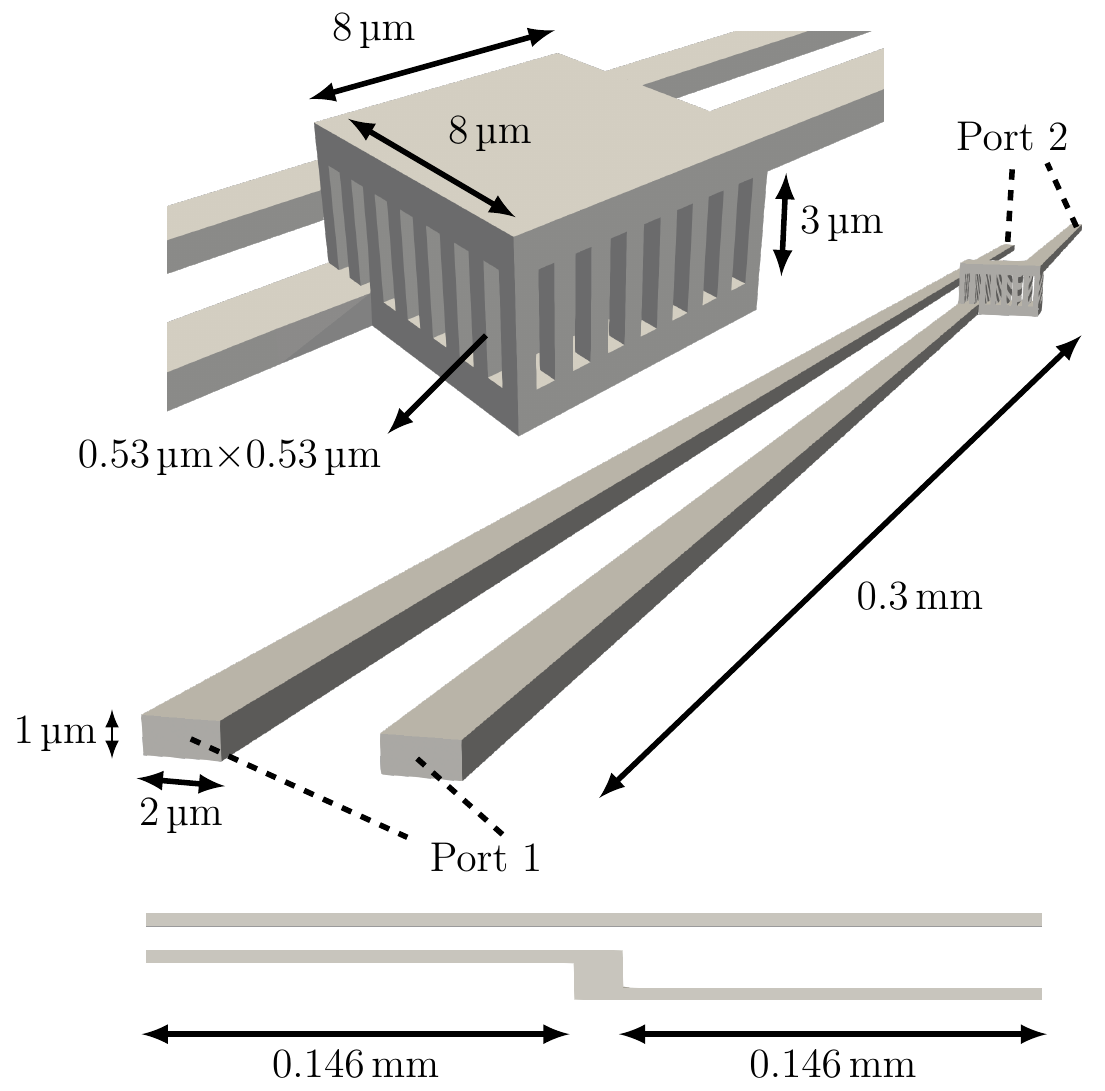}
	\caption{Geometry and port definitions for the differential pair with a microvia array in \secref{sec:results:microvia}.}\label{fig:microvia:geom}
\end{figure}

\begin{figure}[t]
	\centering
	\includegraphics[width=\linewidth]{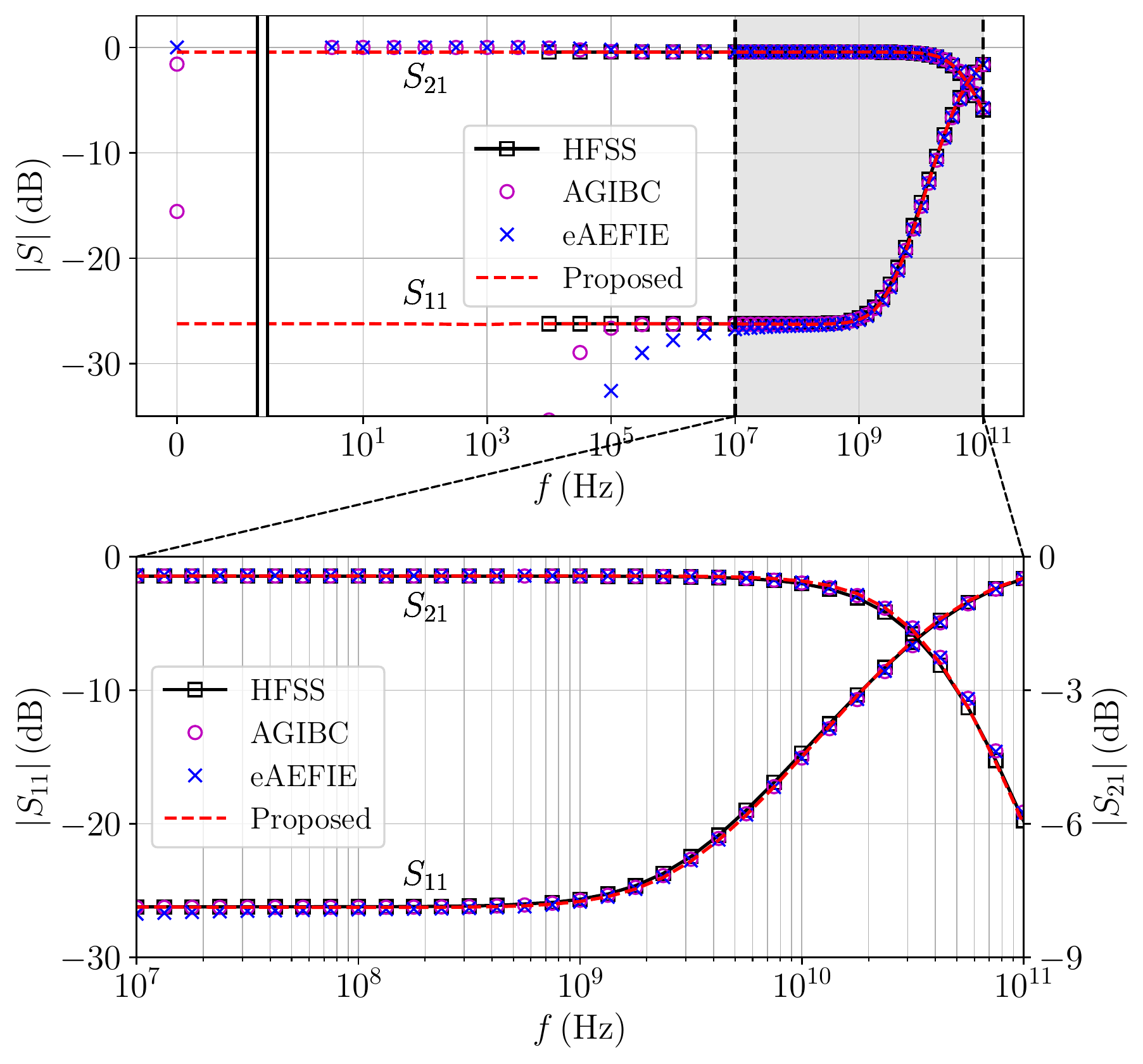}
	\caption{Scattering parameters (magnitude) for the differential pair with a microvia array in \secref{sec:results:microvia}.}\label{fig:microvia:Smag}
\end{figure}

\begin{figure}[t]
	\centering
	\includegraphics[width=\linewidth]{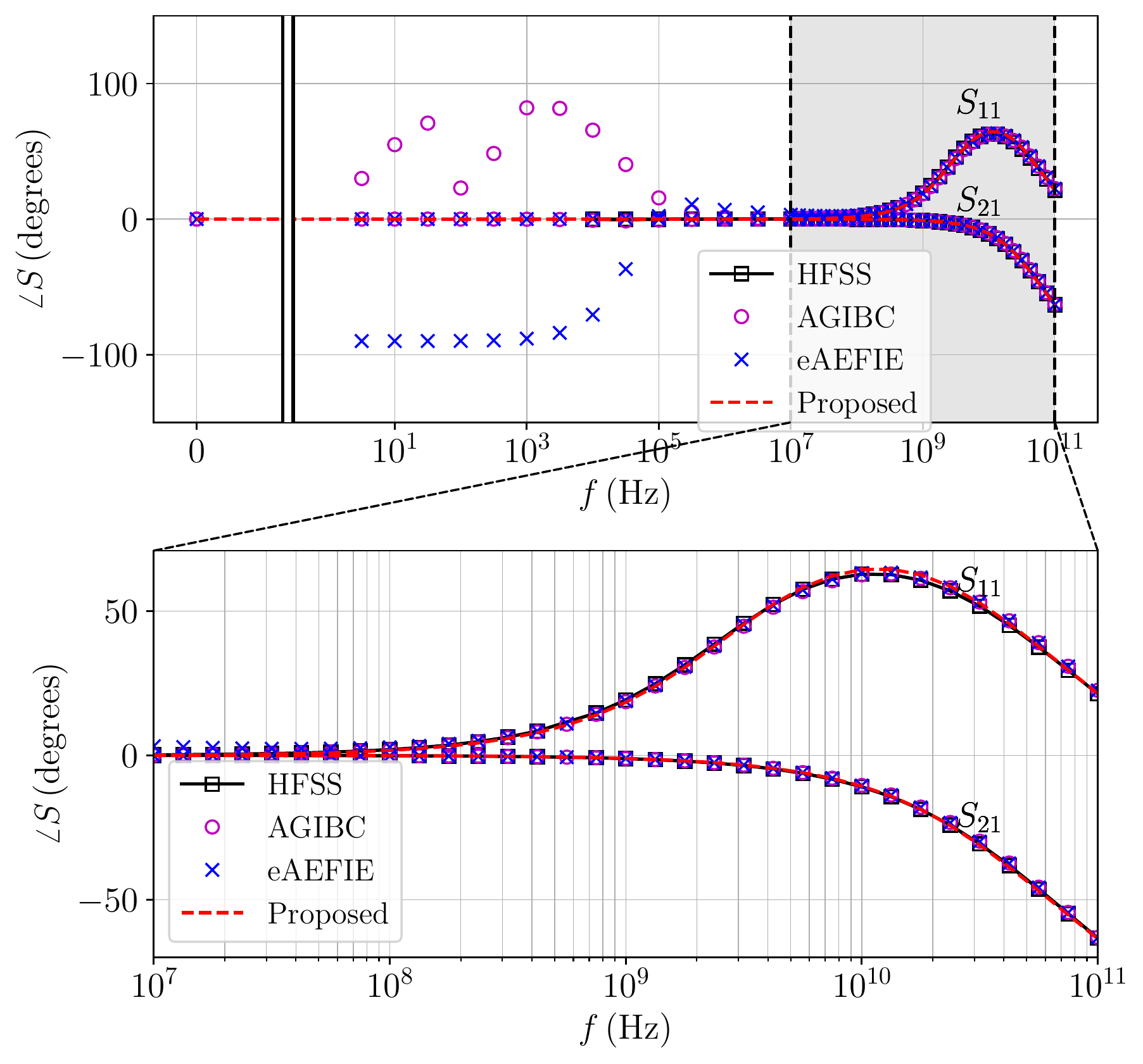}
	\caption{Scattering parameters (phase) for the differential pair with a microvia array in \secref{sec:results:microvia}.}\label{fig:microvia:Sang}
\end{figure}

\begin{figure}[t]
	\centering
	\includegraphics[width=\linewidth]{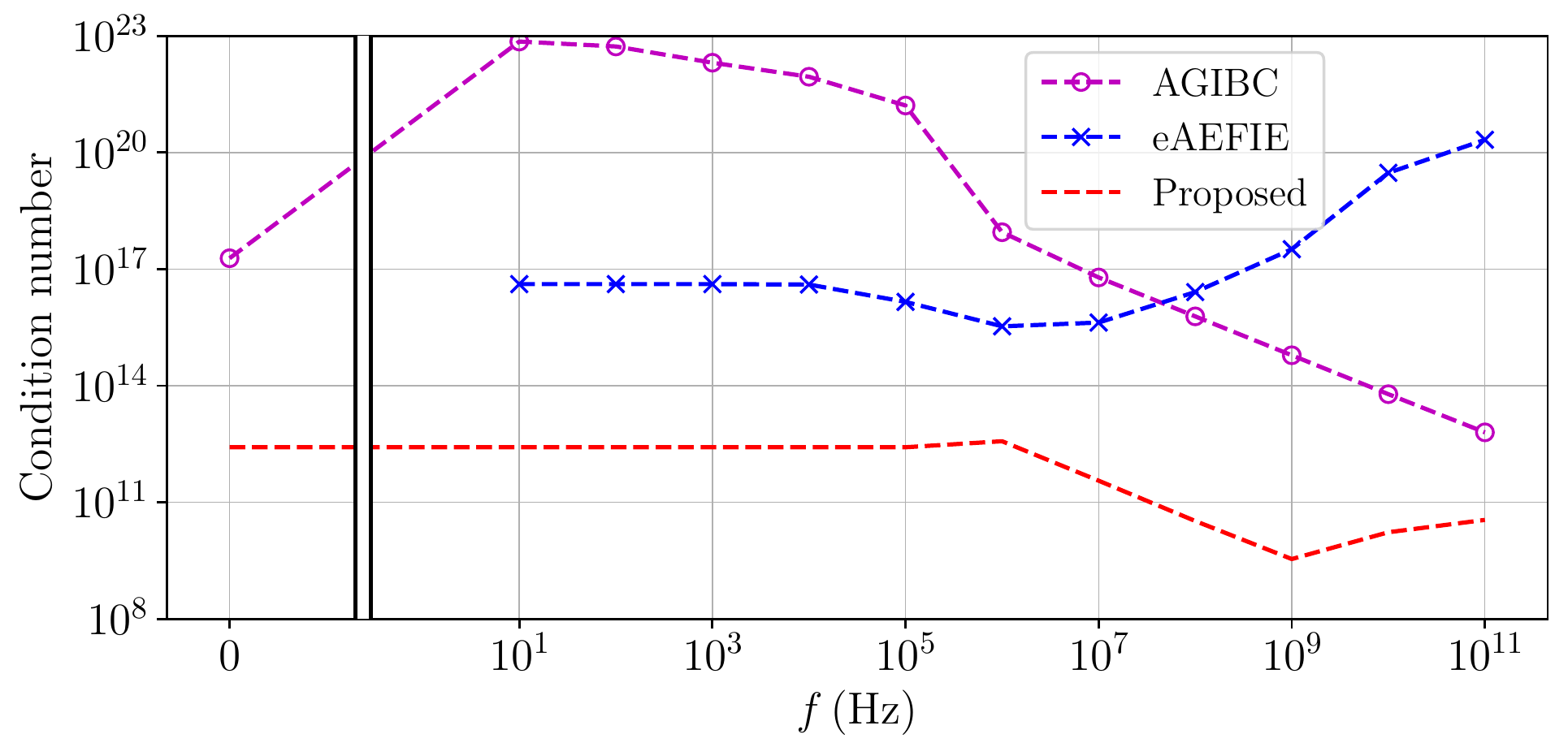}
	\caption{Condition number of the system matrix in~\eqref{eqd:sys2} for the differential pair with a microvia array in \secref{sec:results:microvia}, compared to that of field-based formulations.}\label{fig:microvia:cond}
\end{figure}

Next, we consider a differential pair composed of copper traces, where one of the traces transitions to a different elevation through an~$8\times 8$ array of microvias.
This is a challenging structure which contains multiscale features, and the geometry and port definitions are provided in \figref{fig:microvia:geom}.
The structure is meshed with~$3{,}662$ triangles.
Again, the magnitude (\figref{fig:microvia:Smag}) and phase (\figref{fig:microvia:Sang}) of the~$S$ parameters show excellent agreement between the proposed method and all other methods at high frequency.
Again, the proposed approach remains accurate down to DC, while HFSS encounters an error below~${\sim}10\,$kHz and the field-based methods become inaccurate below~${\sim}100\,$kHz.
\figref{fig:microvia:cond} shows the condition number of the system matrix in~\eqref{eqd:sys2} over the entire frequency range, and confirms that the condition number remains constant at very low frequencies, indicating that the proposed formulation is stable down to DC despite the nontrivial geometry.

\subsection{On-Chip Inductor Coil}\label{sec:results:ind}

\begin{figure}[t]
	\centering
	\includegraphics[width=\linewidth]{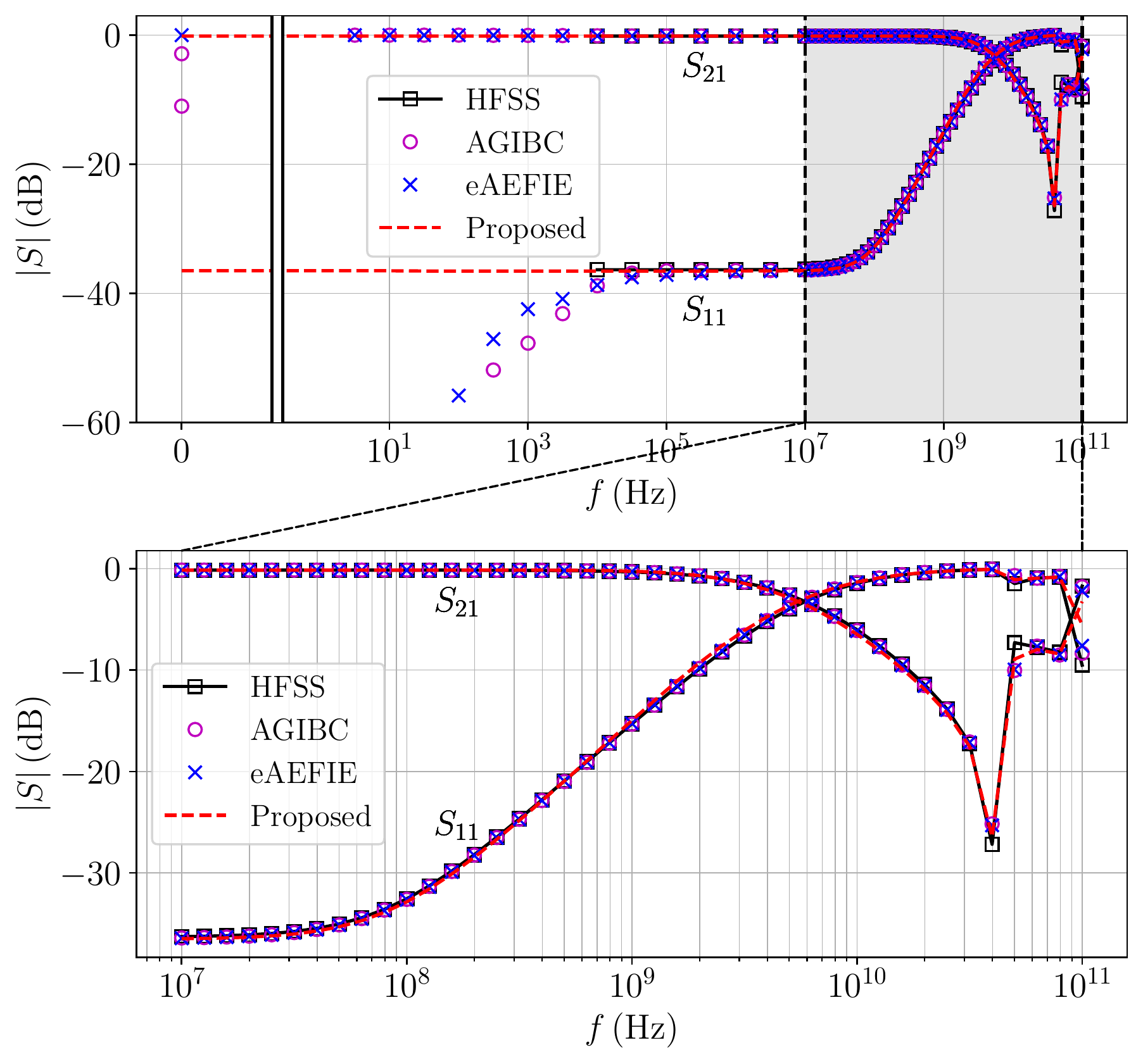}
	\caption{Scattering parameters (magnitude) for the inductor coil in \secref{sec:results:ind}.}\label{fig:ind:Smag}
\end{figure}

\begin{figure}[t]
	\centering
	\includegraphics[width=\linewidth]{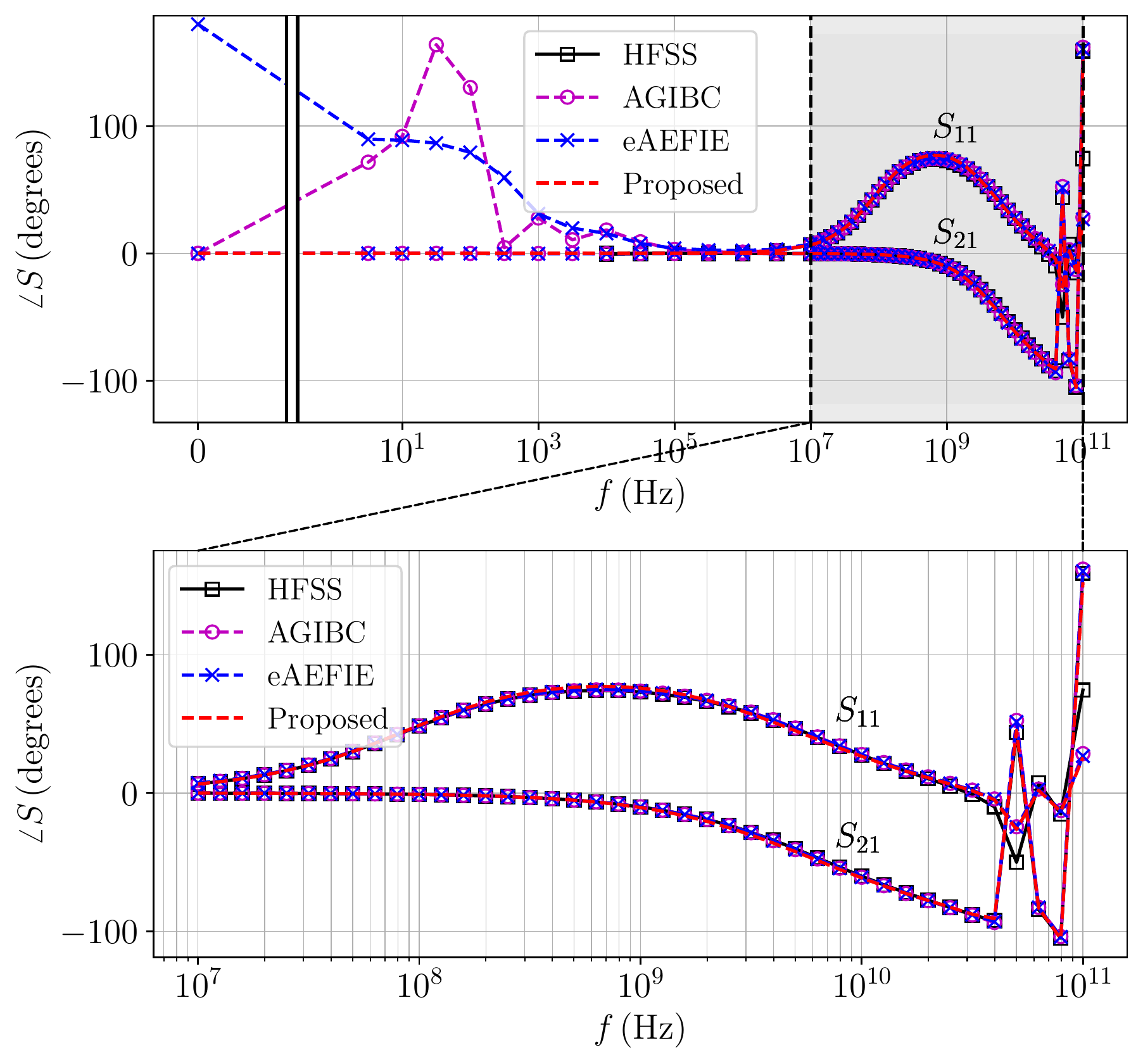}
	\caption{Scattering parameters (phase) for the inductor coil in \secref{sec:results:ind}.}\label{fig:ind:Sang}
\end{figure}

\begin{figure}[t]
	\centering
	\includegraphics[width=\linewidth]{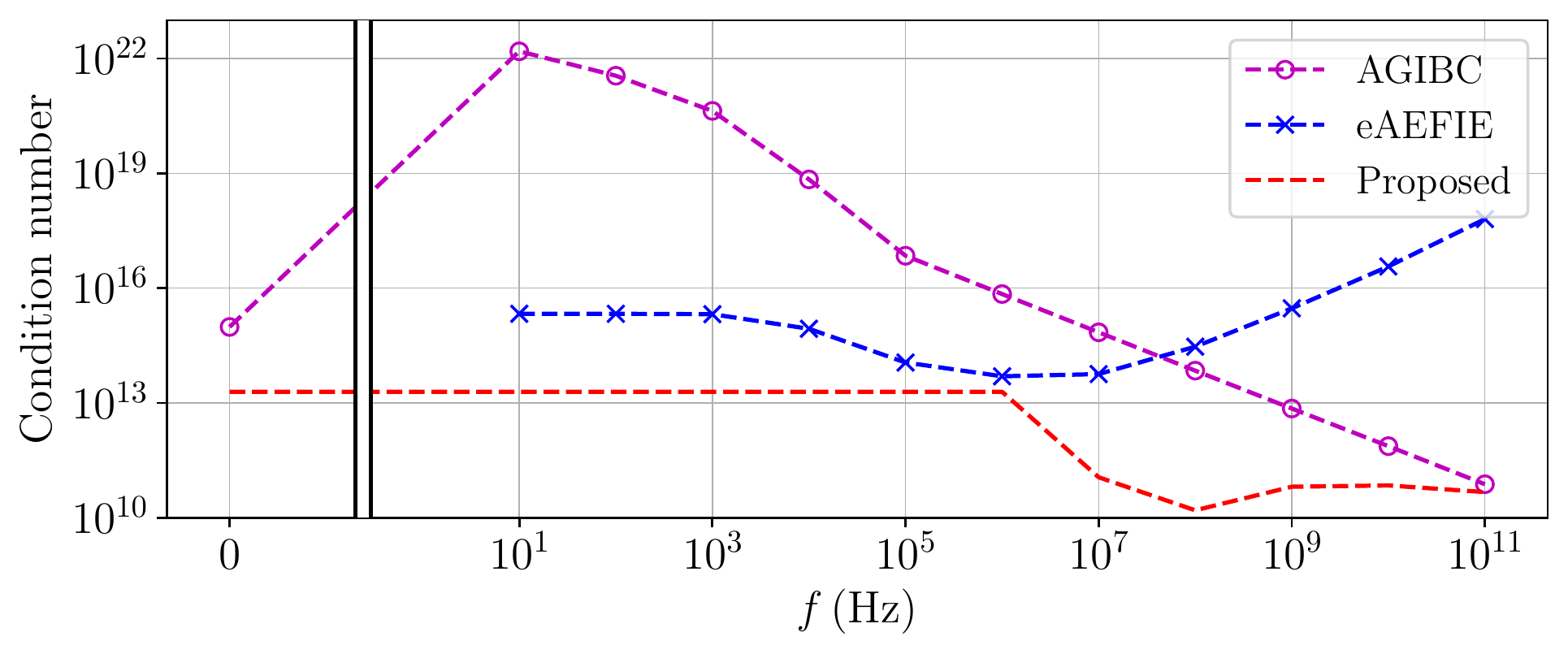}
	\caption{Condition number of the system matrix in~\eqref{eqd:sys2} for the inductor coil in \secref{sec:results:ind}, compared to that of field-based formulations.}\label{fig:ind:cond}
\end{figure}

An on-chip copper inductor coil is considered here, which is a~$4\times$ scaled version of the geometry described in~\cite{EPEPS2017} and is based on the structure in~\cite{fastmaxManual}.
The two-port structure is meshed with~$2{,}030$ triangles.
\figref{fig:ind:Smag} and \figref{fig:ind:Sang} show the excellent accuracy of the proposed method in both magnitude and phase, respectively.
In the high-frequency regime, the proposed method captures the resonance correctly, while at intermediate frequencies, it models the variations in skin depth accurately.
It is also the only technique among the ones considered which remains accurate down to DC as in the previous examples, whereas HFSS and the field-based methods cannot be applied below~${\sim}10\,$kHz.
\figref{fig:ind:cond} demonstrates the stable condition number at very low frequencies, again demonstrating the robustness of the proposed formulation for nontrivial geometries.

\subsection{Part of an IC Package with Vias and Bondwires}\label{sec:results:package}

\begin{figure}[t]
	\centering
	\includegraphics[width=1.0\linewidth]{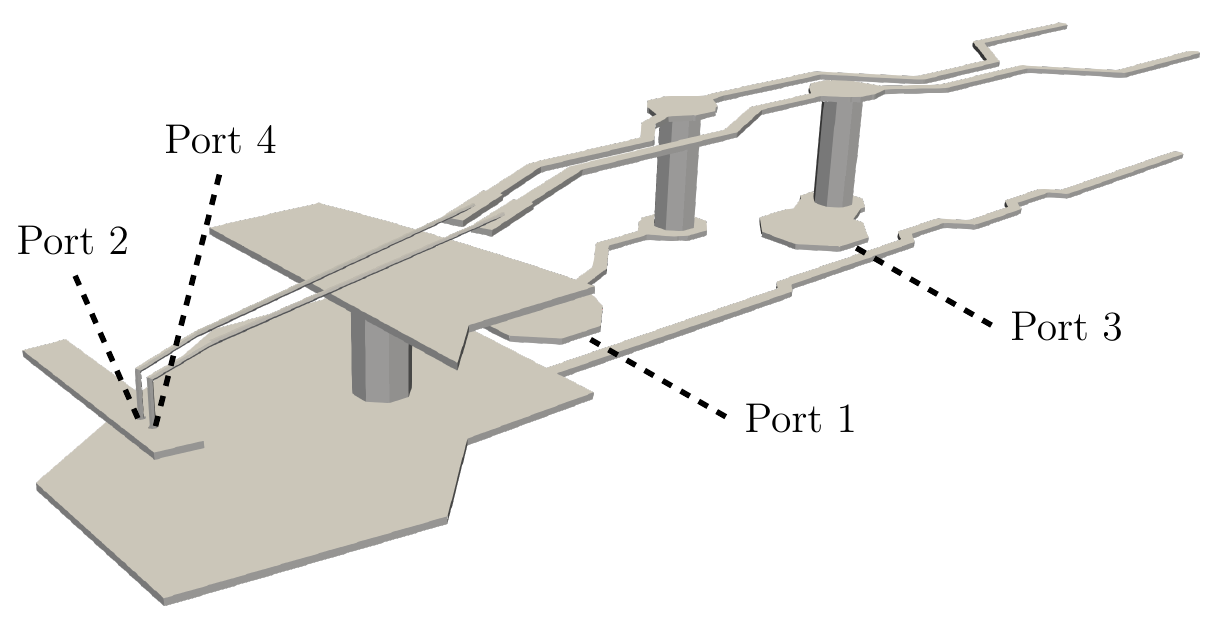}
	\caption{Geometry and port definitions for the part of an IC package in \secref{sec:results:package}.}\label{fig:package:geom}
\end{figure}

\begin{figure}[t]
	\centering
	\includegraphics[width=\linewidth]{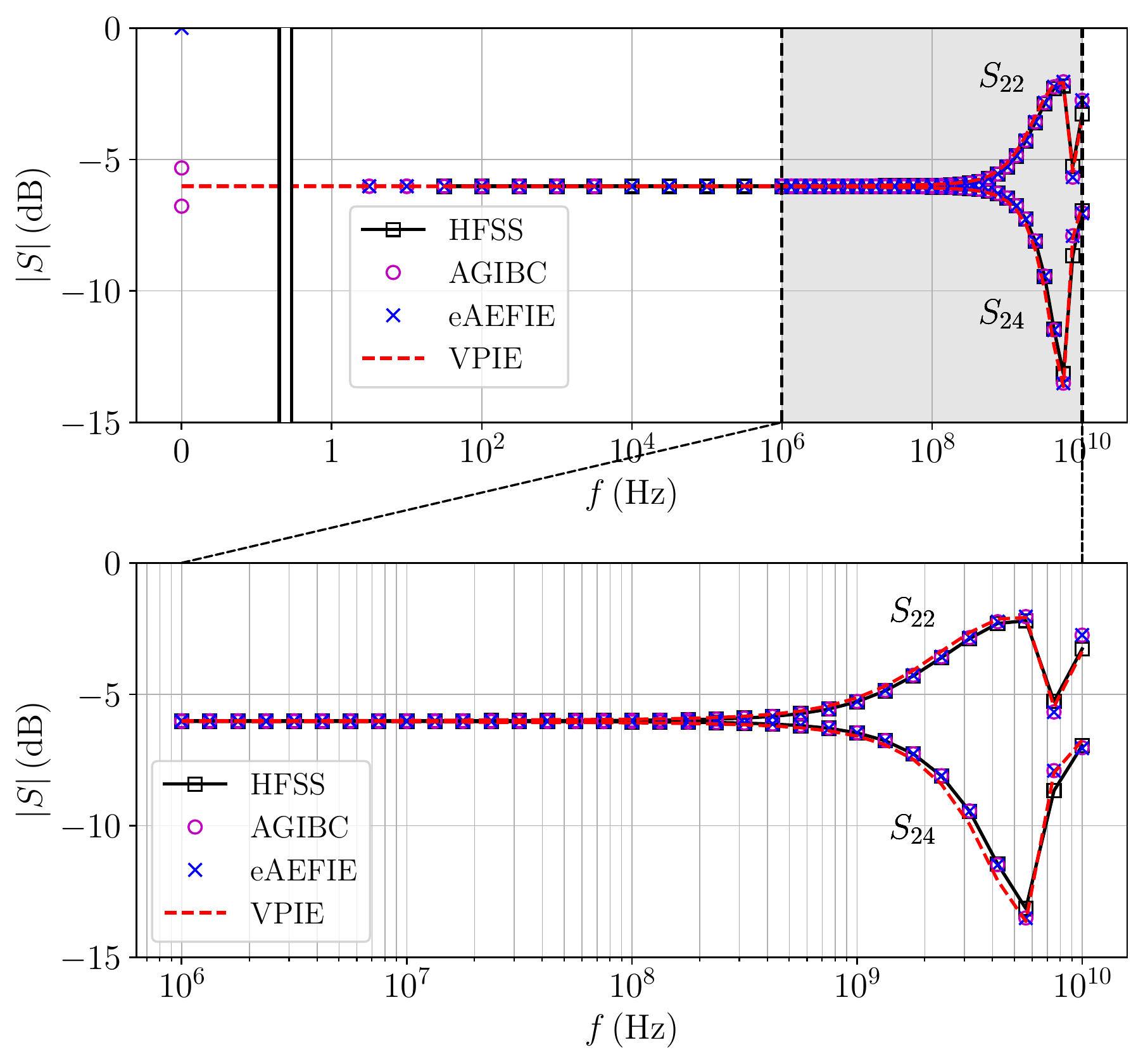}
	\caption{Scattering parameters (magnitude) for the part of an IC package in \secref{sec:results:package}.}\label{fig:package:Smag}
\end{figure}

\begin{figure}[t]
	\centering
	\includegraphics[width=\linewidth]{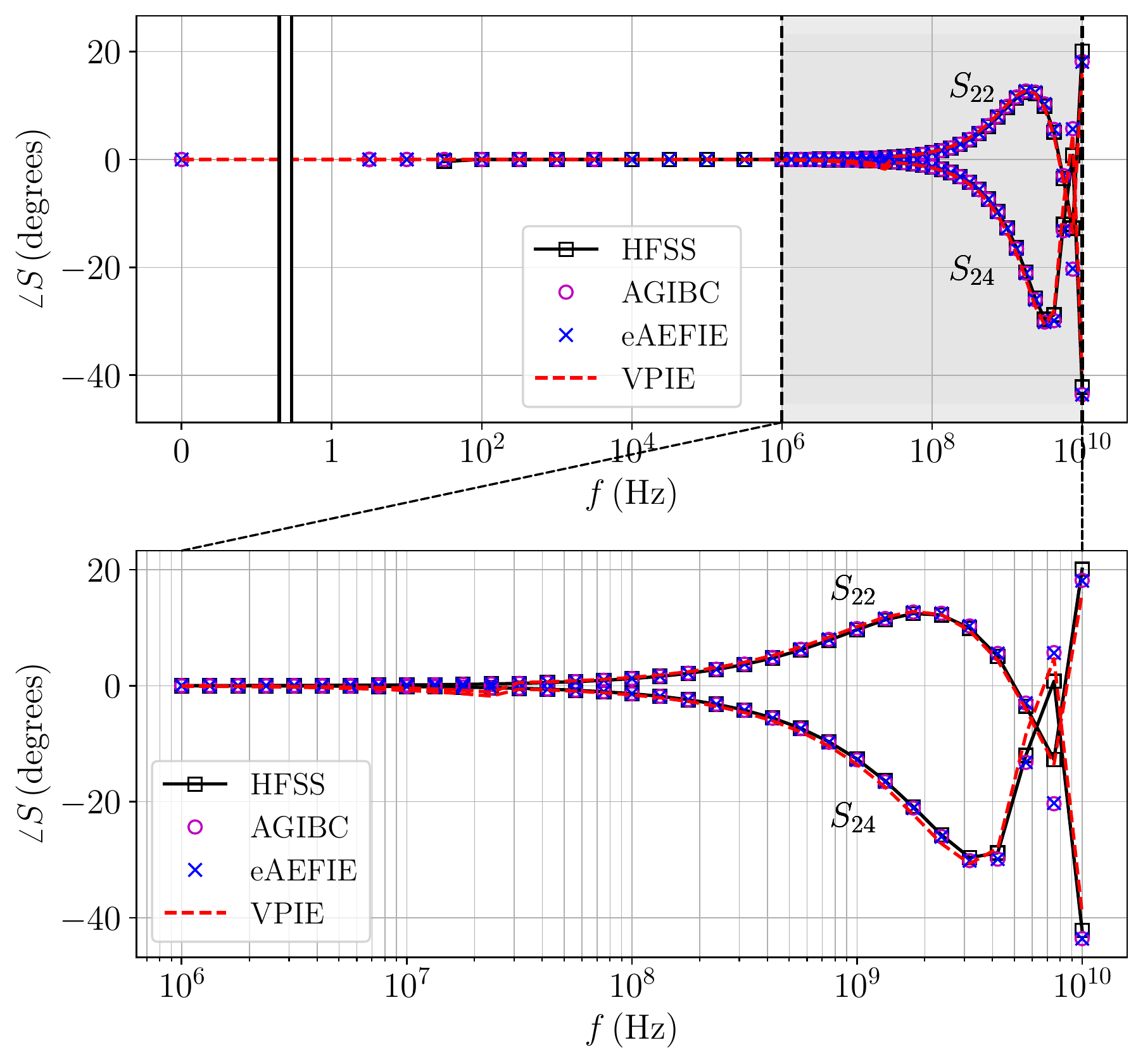}
	\caption{Scattering parameters (phase) for the part of an IC package in \secref{sec:results:package}.}\label{fig:package:Sang}
\end{figure}

Finally, we consider a package-level structure obtained from the library of examples provided by Ansys HFSS.
The structure contains non-trivial features such as vias and bondwires, but was modified to remove the large ground and power planes to ensure that the LU factorization could be performed within the available computational resources.
The geometry and port definitions of the IC package section are shown in \figref{fig:package:geom}; the structure was meshed with~$5{,}058$ triangles.
The proposed method captures the high-frequency response of the structure with excellent accuracy, as confirmed in the bottom panels of \figref{fig:package:Smag} and \figref{fig:package:Sang}.
It is also accurate at DC, unlike the other methods considered.
In this case, HFSS cannot solve the problem below~${\sim}10\,$Hz, while the field-based methods become inaccurate below~${\sim}1\,$Hz.

Although the use of acceleration algorithms, such as the fast multipole method~\cite{FMAorig,enghetaFMA} and those based of fast Fourier transforms~\cite{AIMbles,fastimp}, will be necessary to allow modeling larger structures, the realistic examples considered here demonstrate the excellent accuracy and stability of the proposed method for both chip- and package-level applications.
Its ability to model structures continuously down to DC can be useful in signal integrity analysis, where digital and mixed signals may contain DC components in addition to high frequencies, and for power integrity applications, where a DC supply with minimal voltage ripple is desirable.
The proposed potential-based formulations is therefore a promising alternative to field-based methods for the wideband full-wave modeling of electrical interconnects.

\section{Conclusion}\label{sec:concl}

We have presented a new boundary element formulation based on electromagnetic potentials rather than fields, for the full-wave modeling of electrical interconnects.
A set of vector potential integral equations was devised for modeling the regions both internal and external to each conductive object, taking into account the coupling to an external circuit.
This article systematically addressed several mathematical challenges associated with the use of potentials instead of fields.
Several realistic numerical examples demonstrate that the proposed formulation yields accurate network parameters from DC to very high frequencies, and has a condition number which remains stable as the frequency decreases.
To the best of our knowledge, the ability to model lossy conductors continuously down to DC has not been achieved by existing full-wave BEM formulations.
This article demonstrates that potential-based techniques, such as the one presented here, hold great promise for the broadband analysis of electrical interconnects at the chip and package levels.



\appendix[Existence of~$\chir$ to justify~\eqref{eq:bcAnr} and~\eqref{eq:bcnchir}]\label{sec:app}
To obtain~$\Arin{\mathcal{V}_0}$ via the gauge transformation~\eqref{eq:Atransform} such that~\eqref{eq:bcAnr} is satisfied, a function~$\chir$ must exist which satisfies~\eqref{eq:bcnchir}.
The goal here is to formulate a boundary value problem for~$\chir$ and demonstrate that a solution satisfying~\eqref{eq:bcnchir} always exists.

Using~\eqref{eq:Atransform} in~\eqref{eq:helmAout}, we may write
\begin{multline}
	\nabla^2\,\Apr + k_0^2\,\Apr \\+ \nabla\left(\nabla^2\chir + k_0^2\,\chir\right) = 0,\quad\left(\vect{r}\in\mathcal{V}_0\right).\label{eqa:helmApout}
\end{multline}
To obtain a Helmholtz equation for~$\Apr$ from~\eqref{eqa:helmApout}, we can choose~$\chir$ such that
\begin{align}
	\nabla^2\chir + k_0^2\,\chir = 0,\quad\left(\vect{r}\in\mathcal{V}_0\right).\label{eqa:helmchirout}
\end{align}
Together,~\eqref{eqa:helmchirout} and~\eqref{eq:bcnchir} can be interpreted as an exterior Neumann boundary value problem for the Helmholtz equation~\cite{book:colton},
\begin{subequations}
	\begin{align}
		\nabla^2\chir + k_0^2\,\chir &= 0,\quad\left(\vect{r}\in\mathcal{V}_0\right),\label{eqa:helmchirout2}\\
		\nhat\cdot\nabla\chirin{\mathcal{S}^+} &= -\Anprin{\mathcal{S}^+},\label{eq:bcnchir2}
	\end{align}
\end{subequations}
which has a unique solution~\cite{book:colton}.

Similarly, for~$\vect{r}\in\mathcal{V}$, we can use~\eqref{eq:Atransform} in~\eqref{eq:helmAin} to obtain
\begin{multline}
	\nabla^2\,\Apr + k^2\,\Apr \\+ \nabla\left(\nabla^2\chir + k^2\,\chir\right) = 0,\quad\left(\vect{r}\in\mathcal{V}\right),\label{eqa:helmApin}
\end{multline}
which can be reduced to a Helmholtz equation for~$\Apr$ by requiring
\begin{align}
	\nabla^2\chir + k^2\,\chir = 0,\quad\left(\vect{r}\in\mathcal{V}\right).\label{eqa:helmchirin}
\end{align}
Given a solution~$\chirin{\mathcal{S}^+}$ of~\eqref{eqa:helmchirout2}--\eqref{eq:bcnchir2}, equation~\eqref{eqa:helmApin} together with the boundary condition~\eqref{eq:bcchir} can be interpreted as an interior Dirichlet boundary value problem~\cite{book:colton} for the Helmholtz equation,
\begin{subequations}
	\begin{align}
		\nabla^2\chir + k^2\,\chir &= 0,\quad\left(\vect{r}\in\mathcal{V}\right),\label{eqa:helmchirin2}\\
		\chirin{\mathcal{S}^-} &= \widetilde{\chi}\rin{\mathcal{S}^+},\label{eq:bcchir2}
	\end{align}
\end{subequations}
which also has a unique solution when~$\mathcal{V}$ is conductive and~$\abs{k}>0$~\cite{book:colton}.
Though the solution may not be unique at DC, it is sufficient for our purposes that at least one solution exists~\cite{book:colton}.
The existence of at least one solution of the boundary value problems~\eqref{eqa:helmchirout2}--\eqref{eq:bcnchir2} and~\eqref{eqa:helmchirin2}--\eqref{eq:bcchir2} implies that a function~$\chir$ can always be found such that~\eqref{eq:bcnchir} is satisfied.
This justifies imposing the boundary condition~\eqref{eq:bcAnr} at any frequency.

%

%

%
%

\ifCLASSOPTIONcaptionsoff
  \newpage
\fi



\bibliographystyle{IEEEtran}
\bibliography{./IEEEabrv,./bibliography}
\end{document}